\documentclass[10pt]{amsart}
\usepackage[a4paper, left=0.8in, right=0.8in, top=1in, bottom=1in]{geometry}

\usepackage{fancyhdr}
\usepackage{url}
\usepackage{CJKutf8}
\usepackage[table]{xcolor}
\usepackage{amssymb}
\usepackage{pifont}
\usepackage{pbox}
\usepackage{enumitem}
\usepackage[T1]{fontenc}
\usepackage[utf8]{inputenc}

\usepackage{xcolor}         

\usepackage{caption,subcaption}
\usepackage{multirow}
\usepackage{tikz}

\usepackage{tipa}
\usepackage{hyperref}
\usepackage{amsmath,amssymb,amsfonts}
\usepackage{algorithmic}
\usepackage{graphicx}
\usepackage{enumitem}
\usepackage{textcomp}
\usepackage{subcaption}
\usepackage{wrapfig}
\usepackage{floatflt}

\usepackage{float} 
\usepackage{booktabs}
\usepackage{tikz}
\usepackage{xcolor}
\usepackage{amssymb}
\usepackage{hyperref}
\usepackage{lipsum}

\usepackage{multirow,bm}
\usepackage{adjustbox}
\usepackage{url}
\usepackage{pifont}

\usepackage{titlesec}

\titleformat{\section}
  {\normalfont\Large\bfseries\raggedright} 
  {\thesection.}{1em}{} 

\titlespacing*{\section}{0pt}{3ex plus 1ex minus .2ex}{2ex}
\titleformat{\subsection}
  {\normalfont\large\bfseries\raggedright}
  {\thesubsection}{1em}{}

\titleformat{\subsection}
  {\normalfont\large\bfseries\raggedright}
  {\thesubsection}{1em}{}

\usepackage{caption}
\begin{document}

\makeatletter
\renewcommand{\@secnumfont}{}
\renewcommand{\MakeUppercase}[1]{#1}
\makeatother

\makeatletter
\renewcommand{\@settitle}{%
  \begin{center}%
    \normalfont\LARGE\bfseries
    \let\MakeUppercase\relax 
    \@title
  \end{center}%
}
\makeatother

\title[Voxlect Benchmark]{\fontsize{15pt}{18pt}\selectfont Voxlect: A Speech Foundation Model Benchmark for Modeling Dialects and Regional Languages Around the Globe}

\author[Tiantian Feng et al.]{
\fontsize{11pt}{11pt}\selectfont
Tiantian Feng\textsuperscript{1}, 
Kevin Huang\textsuperscript{1}, 
Anfeng Xu\textsuperscript{1}, \\
Xuan Shi\textsuperscript{1}, 
Thanathai Lertpetchpun\textsuperscript{1}, 
Jihwan Lee\textsuperscript{1}, \\
Yoonjeong Lee\textsuperscript{1}, 
Dani Byrd\textsuperscript{1}, 
Shrikanth Narayanan\textsuperscript{1}
}

\thanks{\textsuperscript{1} University of Southern California, Los Angeles, CA, USA (corresponding emails: tiantiaf@usc.edu.}

\maketitle

\begin{abstract}
\fontsize{11pt}{13pt}\selectfont We present \texttt{Voxlect}, a novel benchmark for modeling dialects and regional languages worldwide using speech foundation models. Specifically, we report comprehensive benchmark evaluations on dialects and regional language varieties in English, Arabic, Mandarin and Cantonese, Tibetan, Indic languages, Thai, Spanish, French, German, Brazilian Portuguese, and Italian. Our study used over 2 million training utterances from 30 publicly available speech corpora that are provided with dialectal information. We evaluate the performance of several widely used speech foundation models in classifying speech dialects. We assess the robustness of the dialectal models under noisy conditions and present an error analysis that highlights modeling results aligned with geographic continuity.
In addition to benchmarking dialect classification, we demonstrate several downstream applications enabled by \texttt{Voxlect}. Specifically, we show that \texttt{Voxlect} can be applied to augment existing speech recognition datasets with dialect information, enabling a more detailed analysis of ASR performance across dialectal variations.
\texttt{Voxlect} is also used as a tool to evaluate the performance of speech generation systems. \texttt{Voxlect} is publicly available with the license of the RAIL family at: \href{https://github.com/tiantiaf0627/voxlect}{https://github.com/tiantiaf0627/voxlect}.

\end{abstract}


\section{Introduction}
\label{sec:introduction}

A ``dialect'' is defined as a variety of a language spoken by a particular regional and/or social group. Specifically, dialects often differ from their standard language in terms of pronunciation, grammar, and vocabulary, whereas the more commonly used term ``accent'' typically refers to differences in pronunciation and prosody (intonation, rhythm, phrasing). In this paper, we primarily focus on dialect classification. For example, the English language has multiple dialects worldwide, including American, British, Singaporean, and Indian English, each with its distinct linguistic features. 
A common example is the word for a mechanical lift, where ``elevator'' is used in American English, while ``lift'' is preferred in British English. 
Similarly, Mandarin Chinese has many regional dialects across China, such as the Beijing and the Sichuan dialects.
For example, while Beijing Mandarin articulates the retroflex consonants such as ``zh \textipa{[\texttoptiebar{\:t\:s}]}'' in the retroflex place of articulation, the Sichuan dialects typically merge them with their alveolar counterparts, pronouncing them as ``z \textipa{[\texttoptiebar{ts}]}''.
Apart from dialectal variation within a single language, some countries, such as India, have a wide range of regional languages (as well as their dialectal varieties), with neighboring states speaking distinct but culturally connected languages, such as Tamil, Telugu, and Malayalam.

Classifying dialects (as well as regional languages) is important for building robust speech technologies that accommodate diverse linguistic contexts. However, automatic speech recognition (ASR) systems often exhibit substantial disparities across dialectal varieties of the same language. For example, while many ASR systems perform well on speech samples of widely explored varieties, their accuracy drops significantly for under-represented dialects such as African American Vernacular English and Chicano English, reflecting biases in their training corpora~\cite{harris2024modeling,chang2024self}. 
Similarly, KeSpeech~\cite{tang2021kespeech} demonstrates that the state-of-the-art ASR systems show significantly lower performance on eight Mandarin dialectal varieties compared to Standard Mandarin.
Such performance differences can reduce the reliability and usability of the systems and their applications, such as virtual assistants for speakers of under-resourced dialects.
By explicitly modeling and recognizing varying dialects, we can not only have a better understanding of the limitations of current speech technologies but also advance the development of more reliable and robust language technologies.

The current literature on modeling speaker dialects has largely focused on English varieties. For example, the Edinburgh International Accents of English Corpus (EdAcc)~\cite{sanabria2023edinburgh} includes English speech from participants with a variety of first language (L1) backgrounds, such as L1-Indian languages. Moreover, the British Isles Speaker Corpus\cite{demirsahin2020open} provides high-quality audio recordings of English utterances from speakers across the British Isles, such as Scotland, Wales, Northern Ireland, and Ireland. 
In contrast, research on modeling speaker dialects in non-English languages remains relatively limited. Much of the existing work in this area focuses instead on broader language identification (LID) tasks~\cite{pratap2024scaling, radford2023robust}. One notable effort in this field is CommonVoice~\cite{ardila2020common}, a large-scale multilingual speech dataset that includes self-reported speaker dialect labels. Building on this, CommonAccent~\cite{zuluaga2023commonaccent} presents one of the few benchmarking efforts for speaker dialect classification in three non-English languages (German, Spanish, \& Italian). Nonetheless, its language coverage is still limited, excluding other widely spoken language families such as Chinese and Arabic.

In this paper, we present \texttt{Voxlect}, one of the first benchmarks for classifying dialects and regional languages from multilingual speech data.
Our proposed \texttt{Voxlect} benchmark show unique contributions compared to prior works: (1) Unlike previous studies that focus on a limited set of dialects, \texttt{Voxlect} enables dialect and regional language classification across an extensive list of languages, including English, Mandarin, Indic Languages, Spanish, German, Italian, French, Brazilian Portuguese, Tibetan, and Arabic. (2) To address inconsistencies in dialect labeling in different datasets, \texttt{Voxlect} maps dialect labels into a unified taxonomy for each language, enabling more consistent cross-corpus analysis. (3) We show the broad utility of \texttt{Voxlect} through two applications in ASR performance evaluation and speech generation system assessment, showing that dialect-level distinctions matter in real-world systems. Our extensive experiments confirm that \texttt{Voxlect} yields reliable estimates of speaker dialects across languages spoken worldwide, which presents unique opportunities for data mining, modeling, and knowledge discovery in dialectal speech data.

\section{Related Work}
\label{sec:related_work}

Table~\ref{tab:related_work} summarizes key prior works on speaker dialect modeling. We categorize the existing literature based on whether it focuses on English or non-English languages.

\begin{table}
    \centering
    
    \caption{Comparison of \texttt{Voxlect} with existing literature in modeling speaker dialects or regional languages.}
    \resizebox{0.65\linewidth}{!}{
    \begin{tabular}{lcccccccc}

        \toprule
        \multirow{1}{*}{\textbf{Dataset/Study}} & 
        \multirow{1}{*}{\textbf{Language}} & 
        \multirow{1}{*}{\textbf{Speaker Dialects Covered}}
        \\

        \midrule
        
        GLOBE~\cite{wang2024globe} & \multirow{3}{*}{English}
        & Global English varieties
        \\

        ParaSpeechCaps~\cite{diwan2025scaling} & & 
        Country-level English varieties
        \\

        Vox-Profile~\cite{feng2025vox} & & 
        Global English varieties
        \\
        \midrule
        
        AIShell-3~\cite{shi2021aishell} & 
        \multirow{2}{*}{Mandarin} &Mandarin varieties
        \\

        KeSpeech~~\cite{tang2021kespeech} & &  Mandarin varieties \\

        \midrule
        ADI~\cite{sullivan2023robustness} & Arabic & Arabic varieties \\
        
        \midrule

        \multirow{2}{*}{CommonAccent~\cite{zuluaga2023commonaccent}} & \multirow{2}{*}{Multilingual} & 
        Global English varieties,
        \\

        & & Germany, Italian, Spanish
        \\
        \midrule

        \multirow{5}{*}{\textbf{Voxlect (Ours)}} & \multirow{5}{*}{Multilingual} & 
        Global English varieties and 
        \\

        & & 10 non-English dialects (Arabic,
        \\

        & & Mandarin and Cantonese, Tibetan,
        \\

        & & Spanish, German, French, Italian
        \\

        & & Thai, Indic, Brazilian Portuguese
        \\

        \bottomrule

    \end{tabular}
    }
    \label{tab:related_work}
\end{table}

\subsection{Modeling English Dialects} 

We summarize several representative works in modeling English Dialects in Table~\ref{tab:related_work}. In particular, CommonAccent~\cite{zuluaga2023commonaccent} introduced a dialect classification benchmark using samples from CommonVoice-en~\cite{ardila2020common}. It reported strong performance in recognizing English accents, such as American, Canadian, and British English, using the wav2vec-xlsr model. GLOBE~\cite{wang2024globe} is a similar effort that develops classifiers using \texttt{HuBERT} pre-trained models on CommonVoice-en to predict similar dialect labels. In addition to directly modeling English-speaking dialects, ParaSpeechCaps~\cite{diwan2025scaling} uses language models to process Wikipedia pages to augment the country-level dialect information associated with each celebrity speaker. Finally, our prior benchmark, \texttt{Vox-Profile}~\cite{feng2025vox}, explores speaker dialect classification by unifying English dialect labels from more than ten datasets. This benchmark provides high-performing English-speaking dialect classification models using Whisper Families~\cite{radford2023robust} and WavLM~\cite{chen2022wavlm}, enabling robust English dialect recognition across diverse speaker samples.

\subsection{Modeling Non-English Dialects} 

Table~\ref{tab:related_work} presents related works on modeling speaker dialects in non-English languages. There has been growing interest in dialect modeling for Mandarin. For example, AIShell-3~\cite{shi2021aishell} provides over 80 hours of multi-speaker Mandarin speech, annotated with regional accents. More recently, KeSpeech~\cite{tang2021kespeech} introduced a large-scale dataset covering eight major Mandarin subdialects, comprising speech recordings from 27,237 speakers, with a total duration of 1,542 hours.
For Arabic, Sullivan et al.~\cite{sullivan2023robustness} conducted experiments for Arabic dialect classification, exploring model performance across both five major dialect groups and a more fine-grained set of 17 specific dialect labels. In addition to modeling speaker dialects within a single language, CommonAccent~\cite{zuluaga2023commonaccent} reports experiments modeling speaker dialects in several languages, including German, Spanish, and Italian. Compared to these prior efforts, our proposed \texttt{Voxlect} benchmark supports dialect classification across a more extensive list of spoken languages and dialects, creating opportunities to develop robust and reliable speech technologies that accommodate different linguistic backgrounds.

\section{\texttt{Voxlect} Benchmark}

\begin{figure}[ht] {
    \centering
    \includegraphics[width=\linewidth]{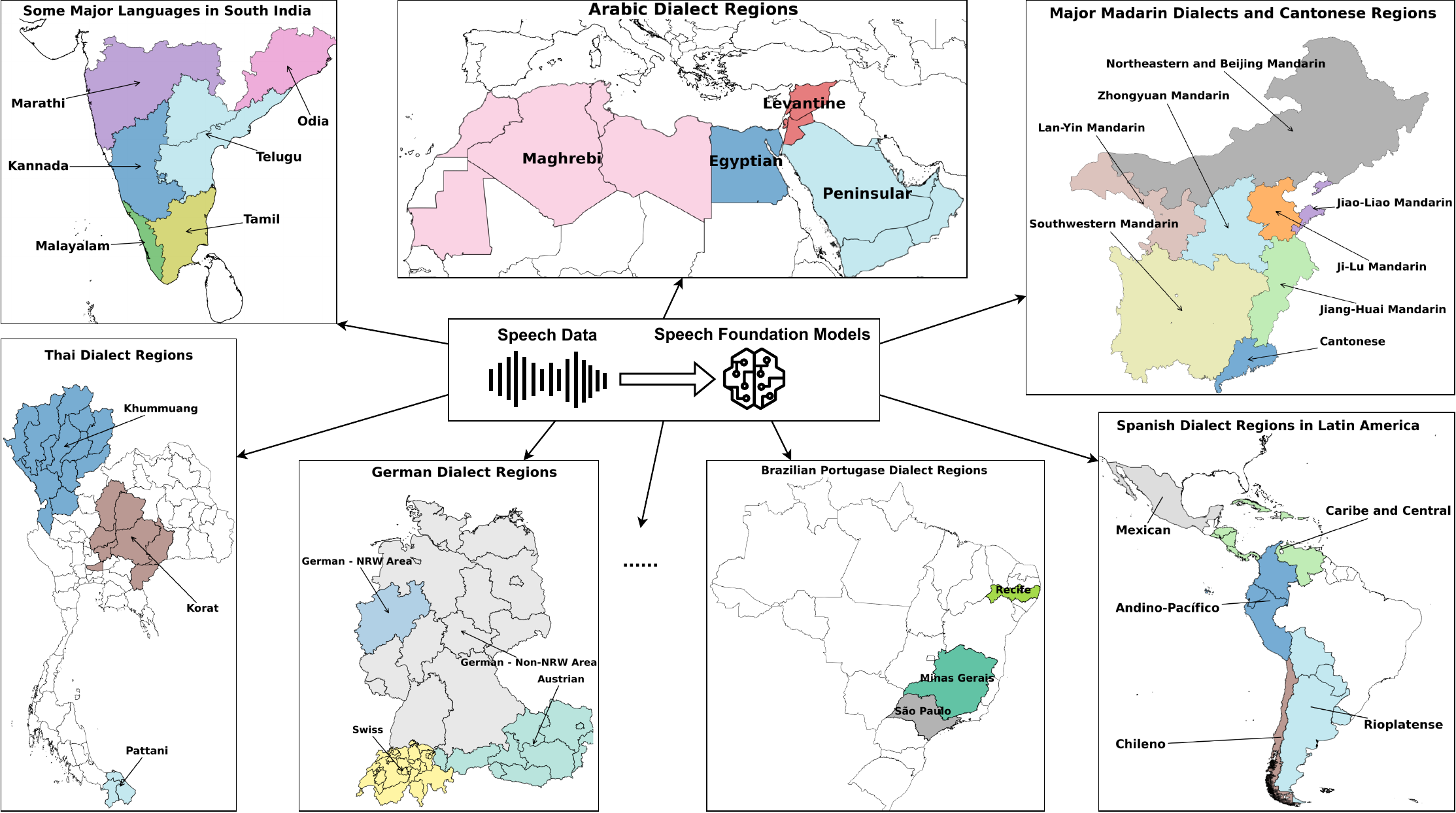}
    
    \caption{Overview of dialect classification in the \texttt{Voxlect} benchmark. The figure plots the geographic distribution of dialects across several language families, including major Indic languages in Southern India and dialects of Mandarin, Arabic, Thai, Spanish, Brazilian Portuguese, and German. Detailed dialect labels and corresponding datasets are provided in Table~\ref{tab:taxonomy}. (The contours shown in the drawing are only schematic, and precise boundaries of dialects are specified in Table~\ref{tab:taxonomy}.)}
    
    \label{fig:voxlect}
} \end{figure}

In this section, we introduce the design of the \texttt{Voxlect} Benchmark. We begin by describing the dialect labels used for each classification. We then outline the speech foundation models used in our experiments. Overall, Figure~\ref{fig:voxlect}\footnote{Disclaimer: The map visualizations in this work are only for illustration purposes for presenting the broad geographic distribution of dialects. The highlighted parts only correspond to those dialects that may be present in experiments. We acknowledge that a single highlighted area may include other dialects or languages. The maps are based on administrative boundary data from GADM-4.1~\cite{gadm2024}. The authors make no claims regarding the accuracy, completeness, or any interpretations of the geographic data provided, and are not responsible for any implications arising from its use.} presents the overview of the \texttt{Voxlect} Benchmark that uses speech foundation models to classify speaker dialects in languages such as Mandarin, German, and Arabic.

\subsection{Labeling Dialects and Regional Languages}
\label{sec:taxonomy}

We note that some dialect labels (e.g., English) have varied usage across existing datasets, highlighting the need for a standardized dialect labeling scheme within each language. This standardization process allows us to combine different available datasets for training reliable dialect classification models. For most non-English cases, dialect labels are defined based on the available dialectal information in the relevant datasets. Specifically, we propose a knowledge-driven dialect categorization for English, Spanish, and French, while adopting existing conventions to group available dialectal data from established datasets for the remaining dialects and regional languages. We summarize the dialect labels and associated experimental datasets in Table~\ref{tab:taxonomy}.

\begin{table*}
    \small
    \centering

    \caption{The labels of speaker dialect in \texttt{Voxlect} benchmark. The green, blue, and violet colors indicate speaker dialects from North America, the British Isles, and other regions or language backgrounds, respectively. For Indic languages, the color green indicates languages of Hindi, Urdu, and English, which are spoken in many regions across India. Moreover, the green and blue indicate Spanish dialects from Europe and Latin America. Similarly, green indicates French and German dialects spoken within France and Germany, while blue represents dialects spoken outside these two countries.}

    \resizebox{\linewidth}{!}{
        \begin{tabular}{cccccccc}
    
            \toprule
            \multicolumn{1}{c}{\textbf{Language}} & 
            \multicolumn{1}{c}{\textbf{Dialect Labels}} & 
            \multicolumn{1}{c}{\textbf{Datasets}} & 
            \multicolumn{1}{c}{\textbf{\#Train Utterances}}
            \\
            \midrule
    
            \multirow{6}{*}{\textbf{English}} & 
            \textcolor{teal}{North America}, \textcolor{blue}{English, Welsh} & CommonVoice-en~\cite{ardila2020common}; CSLU-FAE~\cite{lander2005cslu}; & \multirow{6}{*}{247,098}
            \\
    
             &
            \textcolor{blue}{Scottish, Northern Irish,} \textcolor{violet}{Germanic} & EdAcc~\cite{sanabria2023edinburgh}; British Isles~\cite{demirsahin2020open};
            \\

             &
            \textcolor{blue}{Irish,} \textcolor{violet}{Germanic, Romance,} & L2-ARCTIC~\cite{zhao2018l2}; TIMIT~\cite{garofolo1993darpa};
            \\

             &
             \textcolor{violet}{Slavic, Semitic, Oceania} & VoxPopuli~\cite{wang2021voxpopuli}; ESLTTS~\cite{wang2024usat}; 
            \\
    
             &
            \textcolor{violet}{South Africa, Southeast Asia} & Fair-Speech~\cite{veliche2024towards}; ParaSpeechCaps~\cite{diwan2025scaling} \\
    
             & \textcolor{violet}{East Asia, South Asia, Other} & 
            Nigerian-English~\cite{nigerian_eng}; Hispanic-English~\cite{hispanic_eng};\\
    
            \midrule
    
            \multirow{2}{*}{\textbf{Arabic}} & Egyptian, Levantine, Maghrebi, Peninsular & MASC~\cite{al2023masc}; SADA~\cite{alharbi2024sada};  & \multirow{2}{*}{341,154}
            \\
    
            & Modern Standard Arabic (MSA) & Dvoice~\cite{benelallamdvoice} & 
            \\
            \midrule
    
            \multirow{5}{*}{\shortstack[c]{\textbf{Mandarin} \\ \textbf{and Cantonese}}} & \textcolor{teal}{Standard/Beijing/Northeastern Mandarin} & & \multirow{5}{*}{544,867}
            \\
    
            & \textcolor{blue}{Ji-Lu Mandarin, Southwestern Mandarin,} & KeSpeech~\cite{tang2021kespeech}; 
            \\
    
            & \textcolor{blue}{Jiang-Huai Mandarin,} & CommonVoice-yue~\cite{ardila2020common};
            \\
    
            & \textcolor{blue}{Lan-Yin Mandarin, Zhongyuan Mandarin,} & CommonVoice-hk~\cite{ardila2020common}
            \\

            & \textcolor{blue}{Jiao-Liao Mandarin}, \textcolor{violet}{Cantonese} &
            \\
            \midrule
    
            \multirow{1}{*}{\textbf{Chinese Tibetan}} & U-Tsang, Kham, Amdo & \multirow{1}{*}{TIBMD~\cite{zhao2020open}} & 29,347 \\
    
            \midrule
    
            & \textcolor{teal}{Hindi, Urdu, English}, Punjabi, Dogri & & \multirow{5}{*}{247,302}
            \\
    
            \textbf{Indic} & Kashmiri, Sanskrit, Assamese, Manipuri & IndicVoices~\cite{javed2024indicvoices};
            \\
    
            \textbf{Languages \&} & Bengali, Odia, Maithili, Santali, Gujarati & CommonVoice-en~\cite{ardila2020common} &
            \\
    
            \textbf{Indian English} & Bodo, Marathi, Nepali, Konkani, Sindhi &
            \\

            & Tamil, Telugu, Kannada, Malayalam &
            \\
    
            \midrule
    
            \multirow{1}{*}{\textbf{Thai}} & Khummuang, Korat, Pattani, Thai-central & \multirow{1}{*}{Thai-Dialect-Corpus~\cite{suwanbandit2023thai}} & 302,087
            \\
    
            \midrule
    
            \multirow{2}{*}{\textbf{Spanish}} & \textcolor{teal}{Penisular}, \textcolor{blue}{Mexican, Chileno, Andino-Pacífico}  & CommonVoice-sp~\cite{ardila2020common}; & \multirow{2}{*}{123,102} 
            \\
    
            & \textcolor{blue}{Central America and the Caribbean, Rioplatense} & Latin American Spanish~\cite{guevara2020crowdsourcing}
            \\
    
            \midrule
    
            \multirow{2}{*}{\textbf{French}} & \textcolor{teal}{France}, \textcolor{blue}{Switzerland/Belgium/Germany} & \multirow{1}{*}{CommonVoice-fr~\cite{ardila2020common};} & \multirow{2}{*}{56,280}
            \\

             & \textcolor{blue}{Africa, Canada} & \multirow{1}{*}{African Accented French}
            \\
    
            \midrule
    
            \multirow{2}{*}{\textbf{German}} & \textcolor{teal}{German-Non-NRW Area, German-NRW Area}, & \multirow{2}{*}{CommonVoice-de~\cite{ardila2020common}} & \multirow{2}{*}{71,158}
            \\

            & \textcolor{blue}{Switzerland, Austria, Other} &
            \\
    
            \midrule
    
            \multirow{2}{*}{\textbf{Italian}} & \multirow{2}{*}{Central, Northern, Southern}  & CommonVoice-it~\cite{ardila2020common}; & \multirow{2}{*}{14,883}
            \\

            & & ITALIC~\cite{koudounas2023italic} & \\

            \midrule
    
            \textbf{Brazilian} & \multirow{2}{*}{Minas Gerais, Recife, São Paulo} & \multirow{2}{*}{CORAA~\cite{candido2023coraa}} & \multirow{2}{*}{26,865}
            \\
    
            \textbf{Portuguese} & & & 
            \\
    
            \bottomrule

        \end{tabular}
    }
    \label{tab:taxonomy}
\end{table*}

\subsubsection*{\textbf{English}}

Existing work on modeling English dialects often comes with inconsistent labeling schemes~\cite{wang2024globe,zuluaga2023commonaccent,zhong2025accentbox}. To address this issue, we adopt the dialect taxonomy proposed in our previous \texttt{Vox-Profile} benchmark~\cite{feng2025vox}. In particular, we first categorize two broad regional categories: North America and the British Isles. Within the British Isles, we further distinguish varieties as English (England), Scottish, Northern Irish, Welsh, and Irish. For regions and language backgrounds outside these two, we first define dialects from Oceania and South Africa as two representative English-speaking regions. Next, following common geographic conventions, we categorize English dialects spoken in Asia into three major regions: East Asia, South Asia, and Southeast Asia. 
These Asian regions contain a significant population of native speakers and dialects of English, such as Indian English and Singaporean English.
On top of that, we categorize English accent groups based on their first language (L1) influence, such as Germanic (e.g., German), Slavic (e.g., Russian), Romance (e.g., Spanish, French), and Semitic (e.g., Arabic, Hebrew) language backgrounds. L1 backgrounds with limited data samples in existing datasets are grouped into a miscellaneous general category labeled ``Other.'' For instance, this includes dialects spoken by individuals with Uralic language backgrounds (e.g., Finnish) or from African regions outside South Africa. Compared to \texttt{Vox-Profile}~\cite{feng2025vox}, \texttt{Voxlect} expands the benchmark by integrating the ParaSpeechCaps~\cite{diwan2025scaling} dataset, adding approximately 90,627 additional speech utterances in conversational contexts to the English dialect classification tasks.

\subsubsection*{\textbf{Arabic}}
Arabic dialect classification has been well-established in the literature~\cite{al2023masc,alharbi2024sada,sullivan2023robustness}. 
However, we were unable to obtain the dataset used for the fine-grained 17-dialect classification in~\cite{sullivan2023robustness}. Therefore, we follow conventions in datasets MASC~\cite{al2023masc} and SADA~\cite{alharbi2024sada} and categorize Arabic into five major groups: Egyptian, Levantine (e.g., Lebanon), Peninsular (e.g., Saudi Arabia), Maghrebi (e.g., Morocco), and Modern Standard Arabic (MSA).

\subsubsection*{\textbf{Mandarin and Cantonese}} The classification of Mandarin dialects has been deeply explored within Chinese linguistics. We follow the Mandarin dialect labels listed by KeSpeech~\cite{tang2021kespeech}. Given that there are limited speech samples for Beijing Mandarin and Northeastern Mandarin in KeSpeech and that these two dialects share mutual intelligibility with Standard Mandarin, we group these three varieties into a single category labeled simply Mandarin. However, we highlight that Beijing Mandarin and Northeastern Mandarin do differ from Standard Mandarin in certain lexical items and pronunciations. In addition to classifying Mandarin dialects, we add Cantonese samples from CommonVoice-yue~\cite{ardila2020common} and CommonVoice-hk~\cite{ardila2020common} to enrich the coverage of Chinese languages.

\subsubsection*{\textbf{Chinese Tibetan}} Tibetan is a language spoken across parts of China, India, Nepal, and Bhutan. Since there are limited datasets of Tibetan dialects, we classify three major dialects of Greater Tibet in China as presented in TIBMD~\cite{zhao2020open}: Ü-Tsang, Kham, and Amdo.

\subsubsection*{\textbf{Indic Languages and Indian English}} Many languages, from a number of language families, are spoken in India, and each is typically considered a separate language rather than a dialect. In this paper, we consider 22 major Indic languages (shown in Table~\ref{tab:taxonomy}) as listed in IndicVoices~\cite{javed2024indicvoices}. Moreover, given the widespread use of English across India and its status as an official language of the country, we specifically include Indian English (labels associated with India in CommonVoice-en) as a separate language category to reflect its presence in the multilingual context.

\subsubsection*{\textbf{Thai}} We use the dialect labels described in the Thai-Dialect-Corpus~\cite{suwanbandit2023thai}, one of the most notable efforts to document dialectal variation among Thai speakers. Here, we consider four major dialects: Thai-central, the standard Thai mainly spoken by Central Thai but also across Thailand (e.g., Bangkok); Khammuang, spoken in northern Thailand (e.g., Chiang Mai); Korat, a dialect influenced by both Central Thai and Isan (Northeastern Thai); and Pattani, a Malay-influenced dialect (Southern Thai).

\subsubsection*{\textbf{Spanish}} We categorize Spanish dialects into two broad groups based on geographic landscapes guided by ~\cite{resnick2012phonological}: Peninsular Spanish (spoken in Europe) and Latin American Spanish. Moreover, we break down Latin American Spanish into five major dialectal groups: Mexican, Central America and the Caribbean (e.g., Costa Rica), Chileno (e.g., Chile), Andino-Pacífico (e.g., Peru), and Rioplatense (e.g., Argentina).

\subsubsection*{\textbf{French}} French is spoken by a geographically diverse population worldwide. In addition to its use in France, there is a notable French-speaking population in its neighboring European countries such as parts of Switzerland, Belgium, and Germany. And apart from Europe, French is widely spoken in regions of Africa and Canada (e.g., Quebec). Therefore, we define four French dialect categorizations: France (standard French), the neighboring countries of France (French varieties spoken in Switzerland, Belgium, and Germany), Africa, and Canada. That said, we choose not to refine dialect labels for speakers from France's neighboring countries due to the limited availability of speech samples.

\subsubsection*{\textbf{German}} Akin to our approach to categorizing French dialects, we group German dialects into varieties spoken within Germany and those in other countries with German-speaking populations, adopting the label categories from CommonVoice-de~\cite{ardila2020common}. Having German-North Rhine-Westphalia (NRW) as a separate group owing to its large number of utterances in the dataset, the German dialects are grouped into the following five categories: German-NRW (western Germany), German (non-NRW area), Swiss, Austrian, and the other German-speaking populations.

\subsubsection*{\textbf{Italian}} Given the limited speech datasets presented for Italian dialects and limited speech samples available for individual cities, we create a common regional taxonomy and group Italian dialects into three major categories: Northern Italian (e.g., Venetian), Central Italian (e.g., Tuscan), and Southern Italian (e.g., Sicilian).

\subsubsection*{\textbf{Brazilian Portuguese}} 
There are limited European Portuguese speech datasets with academically friendly licenses, and thus we are unable to include them. In contrast, we rely on CORAA~\cite{candido2023coraa} for studying dialects in Brazilian Portuguese. Based on this dataset, we categorize Brazilian Portuguese into three representative regional groups: São Paulo, Minas Gerais, and Recife.

\begin{figure}[ht] {
    \centering
    \includegraphics[width=0.45\linewidth]{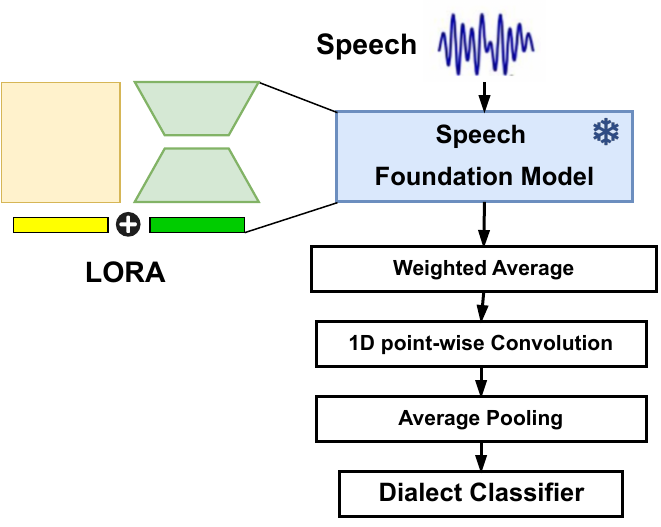}
    
    \caption{Overview of speaker dialect classification architecture in the \texttt{Voxlect} benchmark.}
    
    \label{fig:voxlect_model}
} \end{figure}

\subsection{Speech Foundation Models}
\label{sec:models}

In \texttt{Voxlect}, we evaluate several widely studied speech foundation models, including the Massively Multilingual Speech (\texttt{MMS})~\cite{pratap2024scaling}, \texttt{WavLM}~\cite{hu2024wavllm}, and \texttt{Whisper}~\cite{radford2023robust} family. Among these, MMS and Whisper are multilingual models trained on large-scale cross-lingual datasets, while WavLM is trained only on English. Given the generalizability of these speech foundation models, existing works~\cite{pepino2021emotion, feng2025vox} have shown that direct fine-tuning of hidden outputs from the encoder layers can achieve strong performance in a wide range of downstream speech modeling tasks. We present our modeling architecture in Figure~\ref{fig:voxlect_model}. In our implementation, we first compute a weighted average of the hidden states across all encoder layers, including both convolutional and transformer layers. The aggregated output is then processed through 1D-pointwise convolutional layers. Finally, we average the convolutional outputs to obtain the final embeddings, which are passed through fully connected layers for classification. To further improve the classification performance, we integrate \texttt{LoRa}~\cite{hu2022lora} into all fine-tuning experiments as an effective approach for adapting speech foundation models. 

\section{Experiments}
\label{sec:experiments}

\subsection{Datasets}

We sampled 30 publicly available data sources to conduct benchmark experiments with a total of over 2 million speech samples from the 11 language groups detailed above. For all datasets used in dialect classification experiments, we resample audio to 16 kHz to match the sampling rate of speech foundation models. Audio clips shorter than 3 seconds are excluded, as such short utterances are insufficient for robust dialect classification. To manage computational constraints during training, all samples are truncated to a maximum duration of 15 seconds. For English dialect classification, we discard samples labeled as British, as this lacks specificity on regional varieties such as Scottish. In Spanish dialect experiments, we exclude Colombian speakers from the Latin American Spanish dataset~\cite{guevara2020crowdsourcing}, given the dialectal overlap between the Caribbean and Andino-Pacífico varieties. Due to the large size of the IndicVoice~\cite{javed2024indicvoices} dataset, we subsample a maximum of 10 utterances per speaker. The detailed distribution of dialects or regional languages in each language group is in the Appendix.

\subsection{Experimental Details}

All experiments were conducted using a fixed random seed to ensure reproducibility. During training, we applied several data augmentations to the input waveforms: the Gaussian noise was added with a probability of 1.0, using an SNR range of 3–30 dB; time masking was applied with a probability of 1.0, using a masking ratio between 10\% and 15\%; time stretching was used with a probability of 1.0, with stretch rates ranging from 0.9 to 1.1; and polarity inversion was applied with a probability of 0.5. We use a learning rate of [0.0001, 0.0005] and a training epoch of 15. We perform the training for 5 epochs on Thai and Arabic dialect classification due to a faster convergence rate. Our experiments indicate that models perform better with a learning rate of 0.0005. The pre-trained model weights are downloaded from Huggingface. 

Moreover, we freeze the pre-trained weights in all experiments, while we apply a LoRa with a rank size of 64 to the feedforward layer as suggested by our previous works~\cite{feng2023peft,feng2025vox}. Specifically, we use a batch size of 16 for training the \texttt{Whisper} and \texttt{WavLM} models, and reduce the batch size to 6 for \texttt{MMS-LID-256} due to its larger parameter size. All pre-trained model checkpoints are downloaded from Huggingface. 
For evaluation, we report utterance-level Macro-F1 and accuracy on the test set for each language group. The default test set from each data source is used as the evaluation set; otherwise, we randomly select 20\% of speakers as the test set.

\begin{table*}
    \scriptsize
    \centering

    \caption{Comparison of different speech foundation models in classifying speaker dialects. Overall, the results show that multilingual models including \texttt{Whisper-Large} and \texttt{MMS-LID-256} achieve the overall best performance, while the speech foundation model pre-trained only with English data (\texttt{WavLM}+) shows relatively lower performance in dialect prediction. Bold and underlines indicate the best and the second best performance, respectively.}
    \vspace{-2.2mm}
    \resizebox{0.98\linewidth}{!}{
        \begin{tabular}{lcccccccccccc}
    
            \toprule
             & 
            \multicolumn{2}{c}{\multirow{2}{*}{\textbf{English}}} & 
            \multicolumn{2}{c}{\multirow{2}{*}{\textbf{Arabic}}} & 
            \multicolumn{2}{c}{\textbf{Mandarin}} & 
            \multicolumn{2}{c}{\textbf{Indic Lang and}} & 
            \multicolumn{2}{c}{\multirow{2}{*}{\textbf{Tibetan}}} & 
            \multicolumn{2}{c}{\multirow{2}{*}{\textbf{Thai}}} 
            \\

             & \multicolumn{2}{c}{} & \multicolumn{2}{c}{} & 
            \multicolumn{2}{c}{\textbf{and Cantonese}} & 
            \multicolumn{2}{c}{\textbf{Indian English}} & 
            \multicolumn{2}{c}{} & 
            \multicolumn{2}{c}{} 
            \\
    
            & 
            \multicolumn{1}{c}{\textbf{Acc}} & 
            \multicolumn{1}{c}{\textbf{F1}} & 
            \multicolumn{1}{c}{\textbf{Acc}} & 
            \multicolumn{1}{c}{\textbf{F1}} & 
            \multicolumn{1}{c}{\textbf{Acc}} & 
            \multicolumn{1}{c}{\textbf{F1}} & 
            \multicolumn{1}{c}{\textbf{Acc}} & 
            \multicolumn{1}{c}{\textbf{F1}} & 
            \multicolumn{1}{c}{\textbf{Acc}} & 
            \multicolumn{1}{c}{\textbf{F1}} & 
            \multicolumn{1}{c}{\textbf{Acc}} & 
            \multicolumn{1}{c}{\textbf{F1}} 
            \\
            \cmidrule(lr){1-1} \cmidrule(lr){2-3} \cmidrule(lr){4-5} \cmidrule(lr){6-7}
            \cmidrule(lr){8-9} \cmidrule(lr){10-11} \cmidrule(lr){12-13}
    
            \textbf{Self-Supervised} \\

            \hspace{3mm}\rotatebox[origin=c]{180}{$\Lsh$}WavLM+ &
            79.4 & 0.705 & 
            79.2 & 0.681 &
            79.5 & 0.655 &
            63.8 & 0.634 & 
            80.6 & 0.627 & 
            91.0 & 0.853 
            \\

            \hspace{3mm}\rotatebox[origin=c]{180}{$\Lsh$}MMS-300M &
            65.2 & 0.508 & 
            84.9 & 0.784 &
            78.5 & 0.663 &
            67.6 & 0.688 & 
            79.4 & \underline{0.721} & 
            90.4 & 0.851
            \\

            \hspace{3mm}\rotatebox[origin=c]{180}{$\Lsh$}MMS-LID 256 &
            \underline{80.4} & \underline{0.714} &
            91.8 & 0.860 &
            \textbf{82.9} & \underline{0.708} &
            \textbf{82.6} & \textbf{0.795} &
            \textbf{86.4} & \textbf{0.783} &
            \underline{95.9} & \underline{0.935}
            \\

            \textbf{Whisper Family} \\
            \hspace{3mm}\rotatebox[origin=c]{180}{$\Lsh$}Whisper Tiny & 
            67.2 & 0.545 & 
            88.1 & 0.813 & 
            77.4 & 0.643 & 
            64.6 & 0.642 & 
            79.6 & 0.656 &
            93.2 & 0.896 
            \\

            \hspace{3mm}\rotatebox[origin=c]{180}{$\Lsh$}Whisper Small & 
            78.3 & 0.688 & 
            \underline{93.0} & \underline{0.892} & 
            {81.7} & \textbf{0.712} &
            75.0 & 0.748 &
            78.5 & 0.620 &
            95.4 & 0.931 
            \\
    
            \hspace{3mm}\rotatebox[origin=c]{180}{$\Lsh$}Whisper Large & 
            \textbf{83.0} & \textbf{0.755} & 
            {\textbf{94.2}} & {\textbf{0.923}} & 
            \underline{82.5} & 0.702 & 
            \underline{{77.1}} & \underline{{0.767}} & 
            \underline{82.0} & 0.719 &
            \textbf{96.3} & \textbf{0.943}
            \\

            \midrule

             & &
            \multicolumn{2}{c}{\multirow{2}{*}{\textbf{Spanish}}} & 
            \multicolumn{2}{c}{\multirow{2}{*}{\textbf{French}}} & 
            \multicolumn{2}{c}{\multirow{2}{*}{\textbf{Germany}}} & 
            \multicolumn{2}{c}{\multirow{2}{*}{\textbf{Italian}}} &
            \multicolumn{2}{c}{\textbf{Brazilian}} & 
            \\

             & & \multicolumn{2}{c}{} & \multicolumn{2}{c}{} & 
            \multicolumn{2}{c}{} & 
            \multicolumn{2}{c}{} & 
            \multicolumn{2}{c}{\textbf{Portuguese}}
            \\
    
            & &
            \multicolumn{1}{c}{\textbf{Acc}} & 
            \multicolumn{1}{c}{\textbf{F1}} & 
            \multicolumn{1}{c}{\textbf{Acc}} & 
            \multicolumn{1}{c}{\textbf{F1}} & 
            \multicolumn{1}{c}{\textbf{Acc}} & 
            \multicolumn{1}{c}{\textbf{F1}} & 
            \multicolumn{1}{c}{\textbf{Acc}} & 
            \multicolumn{1}{c}{\textbf{F1}} & 
            \multicolumn{1}{c}{\textbf{Acc}} & 
            \multicolumn{1}{c}{\textbf{F1}} & 
            \\

            \cmidrule(lr){1-1} \cmidrule(lr){3-4} \cmidrule(lr){5-6} \cmidrule(lr){7-8}
            \cmidrule(lr){9-10} \cmidrule(lr){11-12} 
    
            \textbf{Self-Supervised} \\

            \hspace{3mm}\rotatebox[origin=c]{180}{$\Lsh$}WavLM+ &&
            66.2 & 0.662 & 
            59.8 & 0.544 &
            87.6 & 0.769 & 
            64.0 & 0.671 & 
            97.0 & 0.966
            \\

            \hspace{3mm}\rotatebox[origin=c]{180}{$\Lsh$}MMS-300M &&
            58.1 & 0.568 & 
            83.3 & 0.665 &
            89.8 & 0.737 & 
            56.3 & 0.604 & 
            95.1 & 0.942
            \\

            \hspace{3mm}\rotatebox[origin=c]{180}{$\Lsh$}MMS-LID 256 &&
            \underline{77.4} & \underline{0.780} & 
            \underline{86.4} & \underline{0.706} &
            \textbf{96.8} & \textbf{0.906} & 
            \textbf{76.9} & \textbf{0.782} & 
            \textbf{99.1} & \textbf{0.990}
            \\

            \textbf{Whisper Family} \\
            \hspace{3mm}\rotatebox[origin=c]{180}{$\Lsh$}Whisper Tiny && 
            62.5 & 0.630 & 
            72.7 & 0.520 & 
            78.9 & 0.691 & 
            60.2 & 0.614 & 
            86.8 & 0.819
            \\

            \hspace{3mm}\rotatebox[origin=c]{180}{$\Lsh$}Whisper Small && 
            64.3 & 0.650 & 
            83.1 & 0.667 & 
            82.5 & 0.753 & 
            61.9 & 0.622 & 
            94.6 & 0.920
            \\
    
            \hspace{3mm}\rotatebox[origin=c]{180}{$\Lsh$}Whisper Large &&
            \textbf{77.8} & \textbf{0.789} &
            \textbf{87.0} & \textbf{0.712} &
            \underline{93.6} & \underline{0.875} &
            \underline{73.9} & \underline{0.745} &
            \underline{98.6} & \underline{0.980} &
            \\
    
            \bottomrule
    
        \end{tabular}
    }
    \label{tab:voxlect_results}
\end{table*}

\section{Benchmark Results}
\label{sec:results}

\subsection{Dialect Classification Results}

Table~\ref{tab:voxlect_results} presents a comparative analysis of speech foundation models in classifying speaker dialects of 11 language groups. Overall, the models that are pre-trained with multilingual data, particularly Whisper-Large and MMS-LID 256, consistently achieve the best performance in most experiments. Specifically, Whisper-Large achieves the highest accuracy and Macro-F1 in 5 out of 11 language groups, including Arabic (Macro-F1 = 0.923), Mandarin and Cantonese (Macro-F1 = 0.889), and Thai (Macro-F1 = 0.943). On the other hand, \texttt{MMS-LID-256}, a multilingual model fine-tuned for LID, outperforms others in languages like Tibetan (Macro-F1 = 0.783), German (Macro-F1 = 0.906), and Brazilian Portuguese (Macro-F1 = 0.990). In contrast, WavLM+, which is pre-trained only on English data, performs relatively poorly compared to multilingual models, especially across typologically distant languages such as Indic (Macro-F1 = 0.634) and Arabic (Macro-F1 = 0.681). These results highlight the advantages of multilingual models in classifying dialectal variations across different linguistic contexts.

\begin{figure}[ht] {
    \centering
    \includegraphics[width=\linewidth]{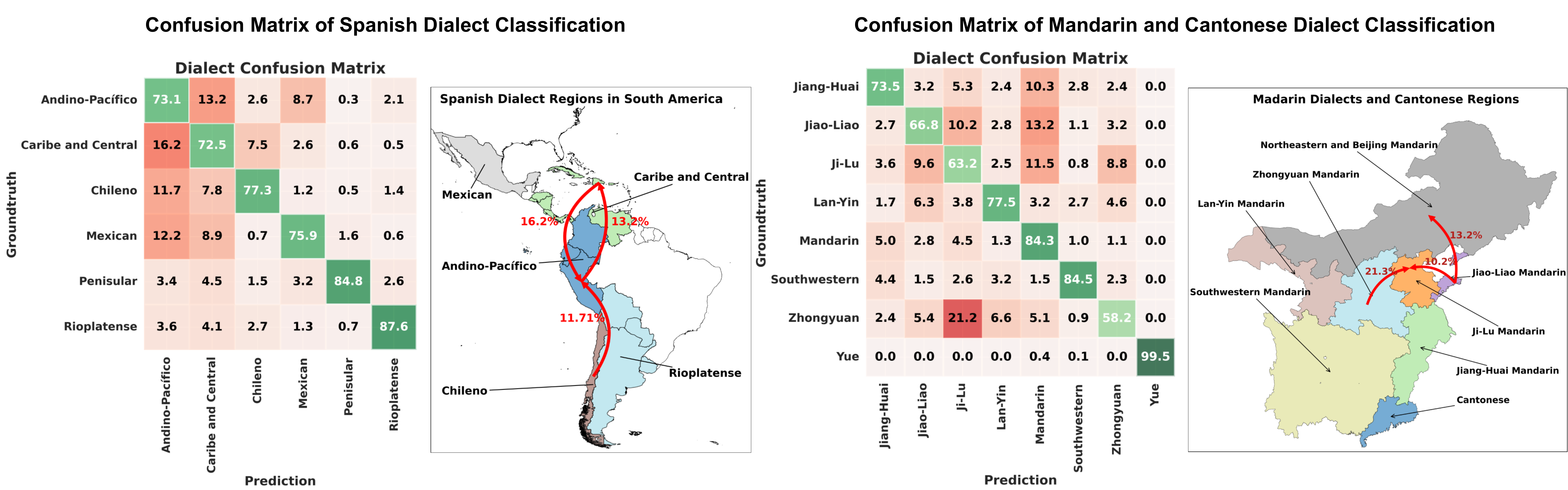}

    \caption{Confusion matrices and geographic visualizations of dialect classification errors for Spanish and Mandarin. The maps highlight the most significant misclassification patterns, with arrows indicating frequently confused dialect pairs. }
    
    \label{fig:map_error}
} \end{figure}

\begin{figure}[h] {
    \centering
    \includegraphics[width=0.55\linewidth]{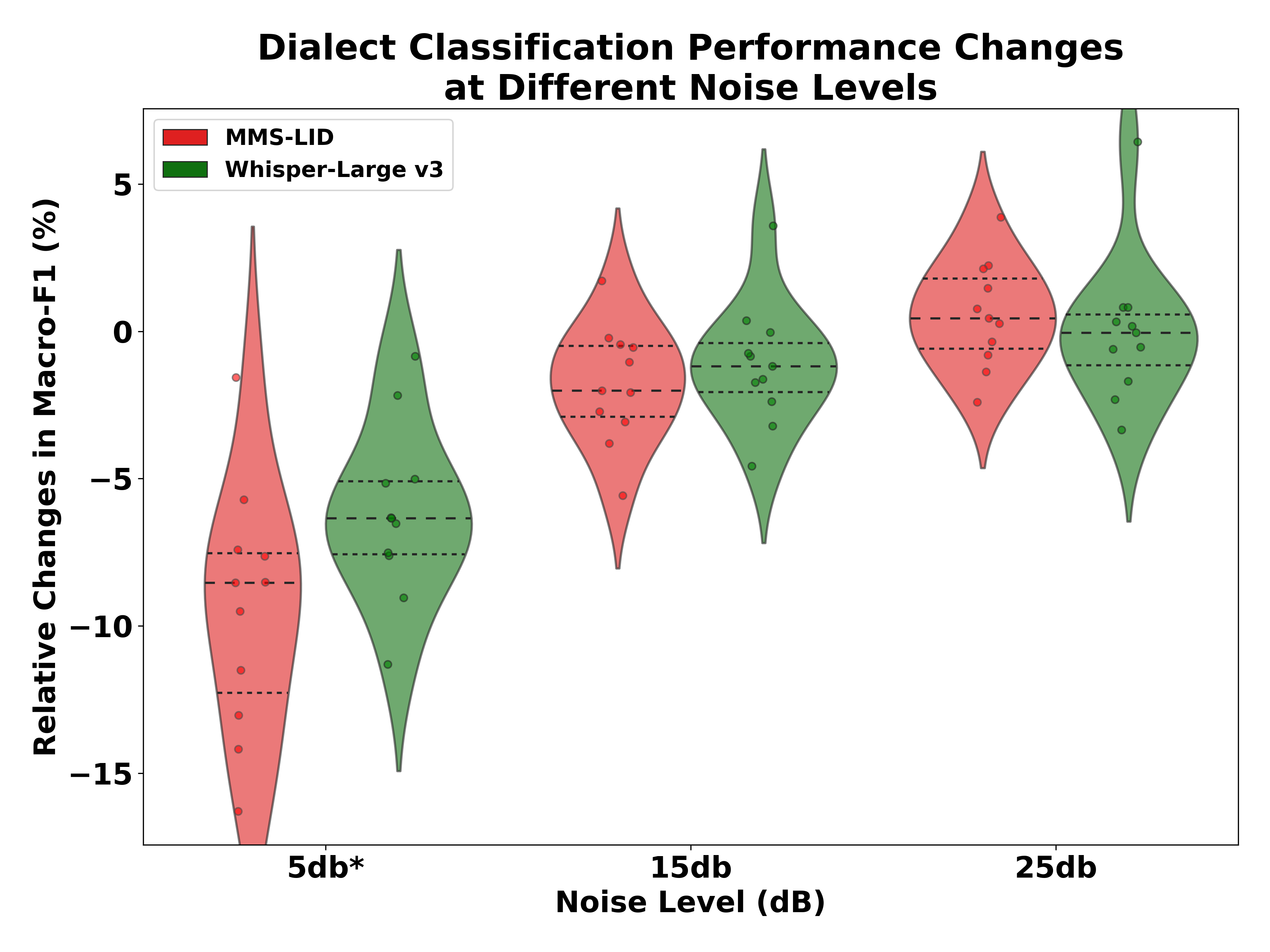}

    \caption{Comparison of relative differences in dialect prediction performance between Whisper-Large and MMS-LID-256 under different noise levels. Each dot represents the performance change for a single dialect classifier of a language, and * indicates statistically significant differences (p$<$0.05).}
    
    \label{fig:noise_robust}
} \end{figure}

\subsection{Geographical Proximity and Classification}

To better understand the performance of these dialect classification models, we visualize both the confusion matrix and the most significant misclassification patterns for two language groups, Spanish and Mandarin, in Figure~\ref{fig:map_error}.
Given that Whisper-Large yields consistently better classification performance than most other models, we create the visualization based on Whisper-Large classification.
Each map highlights the regions of major dialect groups, with arrows and percentages showing the most frequent misclassifications on the evaluation set. 
Overall, we identify a consistent pattern observed in the strong influence of geographic proximity on classification errors. 
Specifically, dialects spoken in neighboring regions or states are more likely to be confused by the dialect classification model. For example, in the Mandarin group, the highest confusion occurs between Zhongyuan and Ji-Lu Mandarin (21.3\%), while in the Spanish group, Caribe and Central dialects are often misclassified as Andino-Pacífico (16.2\%). 
In contrast, dialects that are more geographically distant, such as Peninsular (European) Spanish versus Latin American varieties, or Cantonese versus Mandarin, are less frequently confused. The confusion matrix for each dialect classifier is presented in the Appendix.
These findings suggest that geographical proximity among speakers of varieties of a language often correlates with greater linguistic similarity. While this may present a challenge for a discrete dialect classification, it is consistent with the linguistic study~\cite{chambers1998dialectology,nerbonne2010measuring} of the evolution languages, which recognizes that dialects emerge through contact (and isolation) patterns between peoples over time. 

\subsection{Robustness of Dialect Classification}

We further investigate the robustness of dialect classification under varied noise levels and utterance lengths. The goal here is to assess how robust the fine-tuned models are to real-world scenarios, where acoustic conditions may be degraded and speaking durations vary. 

\subsubsection*{\textbf{Robustness to Noise:}} We introduce the Gaussian noise at three signal-to-noise ratio (SNR) levels: 25dB, 15dB, and 5dB. We compare the relative performance changes of the two best-performing fine-tuned models (Whisper-Large and MMS-LID-256) against their performance on clean speech. The comparisons in Figure~\ref{fig:noise_robust} show that Whisper-Large and MMS-LID-256 demonstrate similar robustness and relatively smaller performance degradation at moderate noise levels (SNR = 15 or 25 dB). However, under the high noise level (SNR = 5 dB), MMS-LID-256 shows a significantly larger drop in dialect prediction performance compared to Whisper-Large.

\subsubsection*{\textbf{Robustness to Utterance Length:}} We compare differences in dialect classification between short and long utterances in Table~\ref{tab:voxlect_utterance}. Specifically, we use a threshold of 6 seconds to define short utterances. The results indicate that, in more than half the cases, the classification performance of both models improves with longer utterances. Particularly, we observe a performance improvement of 0.3 F1 in Indic language classification with longer utterances. 

\vspace{-1mm}
\section{Data-centric Applications of \texttt{Voxlect}}
\label{sec:application}

Here, we demonstrate how \texttt{Voxlect} facilitates two main speech technology applications: analysis of ASR models and automated evaluation of speech generation systems. 

\subsection{Automatic Speech Recognition}

Here, we investigate whether we can leverage the predicted dialect labels to analyze ASR performance across different dialects.
We first predict dialect labels for existing datasets and examine whether these predicted labels lead to the same insights as using the ground truth dialect labels.
Particularly, we predict dialect labels for the test sets of Mandarin and German. For ease of visualization, we include German, Austrian, Swiss, and Other for German, and standard Mandarin, Ji-Lu Mandarin, Southwestern Mandarin, and Zhongyuan Mandarin for Mandarin. We use fine-tuned Whisper-Large models to predict dialects and MMS for ASR.


\begin{table}
    \scriptsize
    \centering

    \caption{Comparison of dialect classification (Macro-F1) between short ($\leq$6 sec) and long (>6 sec) utterances.}
    \resizebox{0.65\linewidth}{!}{
        \begin{tabular}{lcccccccccccc}
    
            \toprule
             & \multicolumn{2}{c}{\textbf{MMS-LID-256}} & \multicolumn{2}{c}{\textbf{Whisper-Large v3}} 
            \\
    
            & 
            \multicolumn{1}{c}{\textbf{Short}} & 
            \multicolumn{1}{c}{\textbf{Long}} & 
            \multicolumn{1}{c}{\textbf{Short}} & 
            \multicolumn{1}{c}{\textbf{Long}} 
            \\
            \cmidrule(lr){1-1} \cmidrule(lr){2-3} \cmidrule(lr){4-5} 
            
            Arabic & 
            \cellcolor{green!10} \textbf{0.856} & 0.842  & 
            0.920 & \cellcolor{green!10} \textbf{0.934}
            \\

            Mandarin & 
            0.707 & \cellcolor{green!10} \textbf{0.710} & 
            \cellcolor{green!10} \textbf{0.702} & 0.701
            \\

            Indic Language & 
            0.690 & \cellcolor{green!10} \textbf{0.977}  & 
            0.659 & \cellcolor{green!10} \textbf{0.971}
            \\

            Tibetan  & 
            0.774 & \cellcolor{green!10} \textbf{0.831} & 
            0.703 & \cellcolor{green!10} \textbf{0.827}
            \\

            Thai  & 
            0.921 & \cellcolor{green!10} \textbf{0.949} & 
            0.921 & \cellcolor{green!10} \textbf{0.966}
            \\

            Spanish  & 
            \cellcolor{green!10} \textbf{0.780} & 0.765 & 
            \cellcolor{green!10} \textbf{0.781} & 0.778
            \\

            German  & 
            0.896 & \cellcolor{green!10} \textbf{0.917} &
            0.858 & \cellcolor{green!10} \textbf{0.892}
            \\

            French  & 
            \cellcolor{green!10} \textbf{0.707} & 0.690 & 
            0.702 & \cellcolor{green!10} \textbf{0.726}
            \\

            \bottomrule
    
        \end{tabular}
    }
    \label{tab:voxlect_utterance}
\end{table}

\begin{figure}[ht] {

    \begin{tikzpicture}
        \node[draw=none,fill=none] at (0, 4.5){\includegraphics[width=0.6\linewidth]{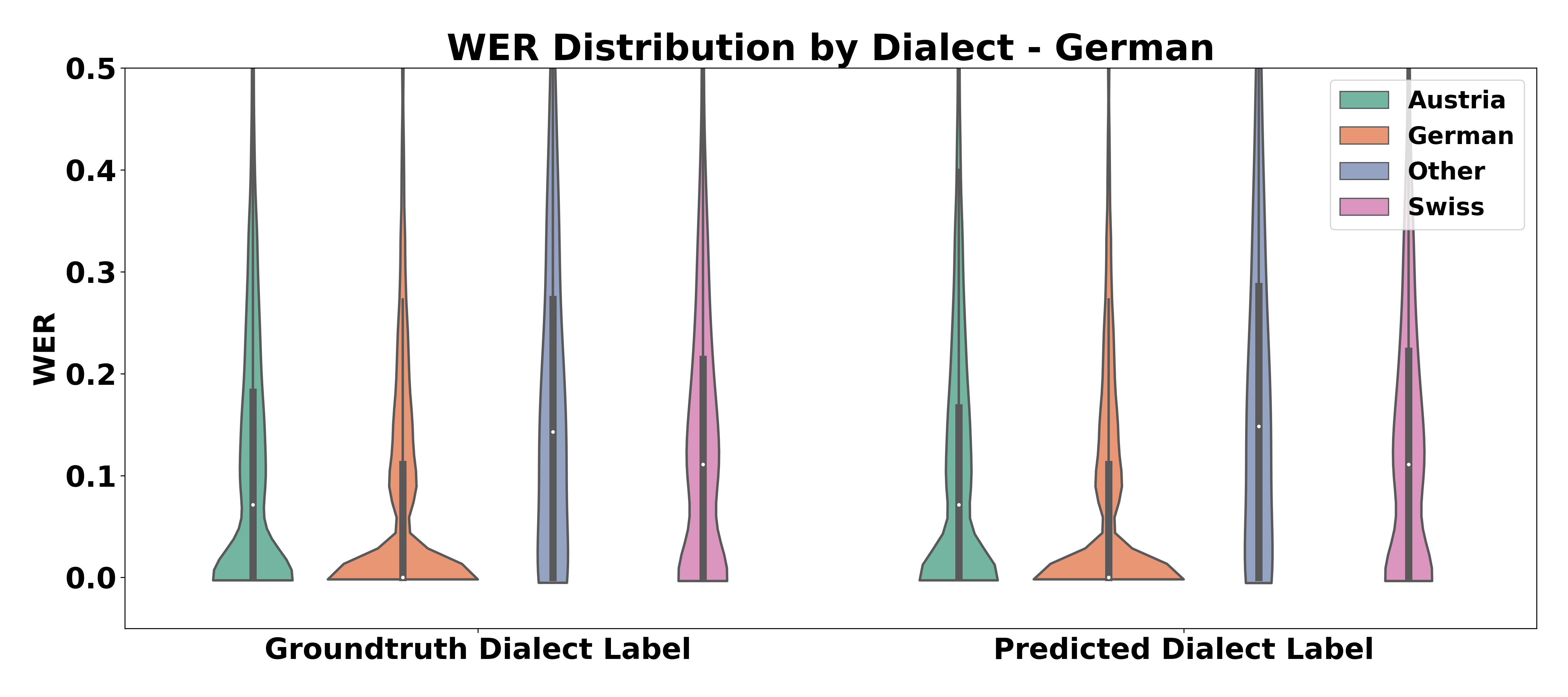}};
        \node[draw=none,fill=none] at (0, 0){\includegraphics[width=0.6\linewidth]{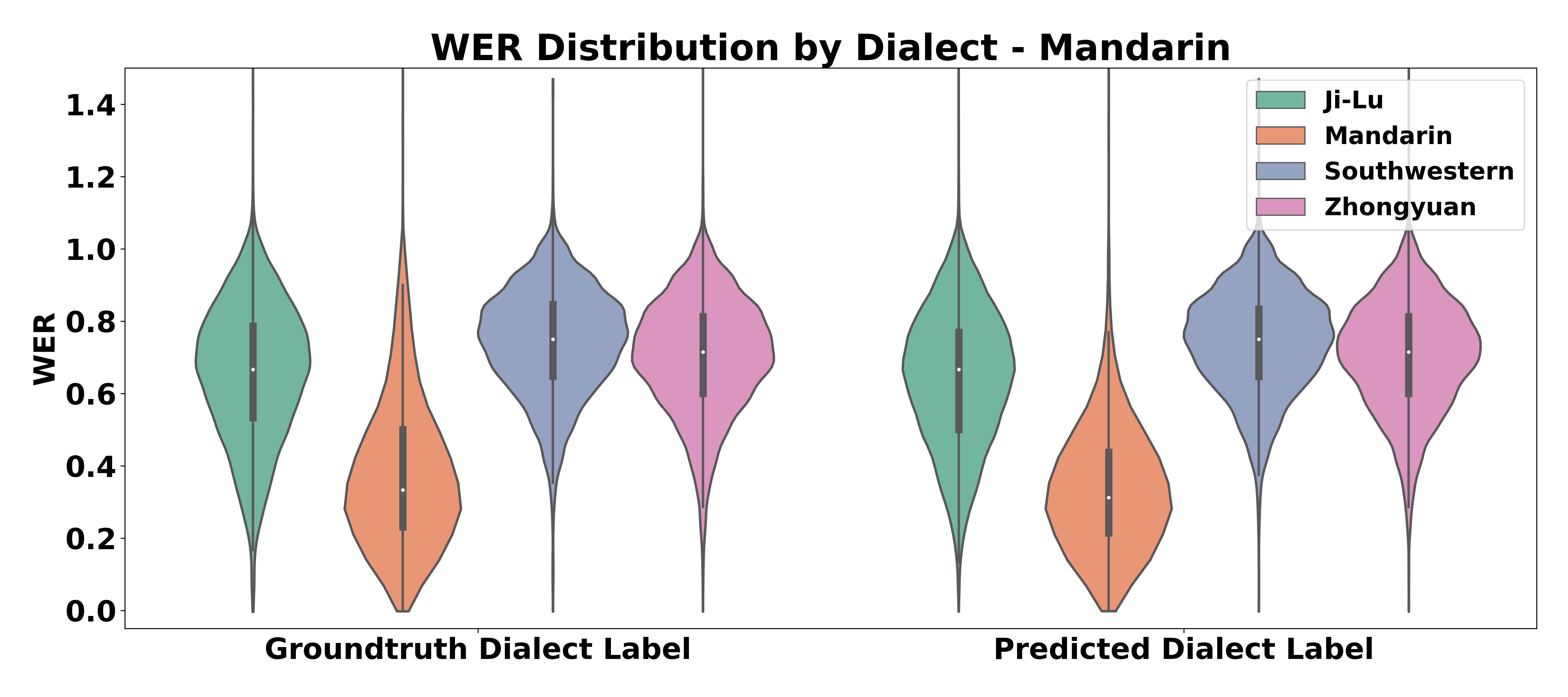}};
    \end{tikzpicture}
    \caption{ASR performance trends, grouped by ground truth and labels predicted by \texttt{Voxlect}.}
    
    \label{fig:voxlect_wer}
} \end{figure}

When computing ASR performance with generated labels, we include only test samples with a predicted dialect probability above 0.7. As shown in Figure~\ref{fig:voxlect_wer}, ASR performance trends using predicted labels from \texttt{Voxlect} closely align with those based on the ground truth labels. In the Mandarin ASR evaluation, speakers of standard Mandarin consistently have lower WER than those using regional sub-dialects, regardless of whether ground truth or predicted labels are used. Among the sub-dialects, Southwestern Mandarin shows the highest WER in both labeling methods. Similarly, we observe that, in the German ASR experiment, speakers labeled as ``German (Non-NRW area)'' demonstrate lower WER compared to those labeled as ``Austria'', ``Swiss'', or ``Other.'' Particularly, the "Other" category shows the highest WER across both ground truth and predicted labels. These findings show that \texttt{Voxlect} provides a reliable tool for identifying limitations in ASR models.

\vspace{-1mm}
\subsection{Evaluation of TTS Systems}

With the rise of media personalization, the technology of generating dialectal speech is popularized in public services and entertainment.
Here, we generate speech samples in specific dialects and investigate whether human evaluations align with evaluations generated by \texttt{Voxlect} regarding the dialect characteristics of the generated speech. Given the availability of publicly accessible TTS models, we focus our experiments on Mandarin. We note that related evaluations for English speech generation are presented in \texttt{Vox-Profile}~\cite{feng2025vox}. Specifically, we use 10 text prompts designed by a phonetician and reference speakers drawn from the KeSpeech test set. To generate speech in five distinct Chinese dialects, we utilize the CosyVoice-2~\cite{du2024cosyvoice}. We use the prompt \begin{CJK*}{UTF8}{gbsn}“用\{方言\}说这句话”\end{CJK*} (which translates to "Use {dialect} to say this sentence"), specifying each of the five dialects: Cantonese, Sichuan (Southwestern), Tianjin (Ji-Lu), Zhengzhou (Zhongyuan), and Shanghai (Jiang-Huai). These were selected to represent a range of major dialectal regions in China. The details of the prompts are in the Appendix.

We invite colleagues with a native language background in these dialects to assess the dialect characteristics of the generated speech samples. Participants were asked to rate the quality of the dialect characteristics on a 5-point scale. We compare the average predicted probability of the corresponding dialects using \texttt{Voxlect} with the averaged human ratings in Table~\ref{tab:voxlect_mos}. The results indicate that the human ratings closely match the model predictions, with the target dialect of Tianjin (Ju-Lu) receiving the lowest ratings and Cantonese receiving the highest scores in both human and machine evaluation.

\begin{table}
    \scriptsize
    \centering

    \caption{Comparison of human evaluation and automated evaluation by \texttt{Voxlect} in assessing quality of dialect characteristics in generated speech of Chinese dialects.}
    \resizebox{0.55\linewidth}{!}{
        \begin{tabular}{lcccccccccccc}
    
            \toprule
             & \multicolumn{1}{c}{\textbf{Human(1-5)}} & \multicolumn{1}{c}{\textbf{\texttt{Voxlect(0-100\%)}}} 
            \\
    
            \cmidrule(lr){1-1} \cmidrule(lr){2-2} \cmidrule(lr){3-3} 
            
            Shanghai(Jiang-Huai) & 
            2.85 & 36.6\%
            \\

            Sichuan(Southwestern) & 
            \cellcolor{green!10} 3.35 & \cellcolor{green!10} 40.4\%
            \\
            
            Tianjin(Ju-Lu) & 
            \cellcolor{red!10} 1.90 & \cellcolor{red!10} 20.5\%
            \\

            Zhengzhou(Zhongyuan) & 
            3.15 & 32.3\%
            \\

            Cantonese & 
            \cellcolor{green!10} 3.50 & \cellcolor{green!10} 53.4\%
            \\
            
            \bottomrule
    
        \end{tabular}
    }
    \label{tab:voxlect_mos}
\end{table}

\section{Limitations and Responsible Use}

While \texttt{Voxlect} introduces a comprehensive benchmark for dialect and regional language classifications, several limitations remain. First, the dialect labels often originate from self-reports, which may contain labeling noise. Second, the benchmark is constrained by the availability of public datasets with existing labels. Thus, refined labels in specific areas (e.g., Paris of France) are not precisely captured, and Mandarin dialects such as those used in Hainan are not studied. Moreover, many globally spoken dialects and languages, such as regional varieties of Korean, Eastern European, or African languages, are currently not included. Third, the robustness of our proposed benchmark has not yet been evaluated beyond different noise levels and utterance lengths, such as cross-domain generalization (e.g., training on read speech and evaluating on natural speech). Finally, dialect classification may potentially expose private data about speakers and raise privacy concerns. While all datasets used in this work are publicly available and \texttt{Voxlect} adheres to the scope of the data usage and license, we take additional measures to mitigate the risk of misuse. Specifically, we release the code and model checkpoints under the Responsible AI License (RAIL) to require responsible use of \texttt{Voxlect}. Users should respect the privacy and consent of the data subjects and adhere to the relevant laws and regulations in their jurisdictions when using \texttt{Voxlect}.

\section{Conclusion and Future Work}
\label{sec:conclusion}

In this work, we propose \texttt{Voxlect}, a benchmark for predicting dialects and regional languages using speech foundation models.
This benchmark includes large-scale machine learning experiments using over 2 million speech utterances from 30 public data sources. 
We experiment with widely adopted speech foundation models and release a suite of high-performing dialects and regional language classification models. 
Benchmarks of this kind in multilingual contexts are rarely represented in the literature.
Our dialect classification results demonstrate that geographic proximity is reflected in dialect similarity. 
While this creates challenges for strict classification, it could bring insight into the cultural and historical factors impacting language evolution. 
Moreover, \texttt{Voxlect} enables a wide range of downstream applications, including the analysis of speech recognition performance and the evaluation of speech generation systems. 
Our next steps include expanding the scope of the benchmark by integrating dialects of additional languages. For example, we plan to include dialectal variations in Korean, which show rich regional diversity such as the Seoul, Gyeongsang, and Jeolla dialects. We also plan to apply our benchmark models to enrich existing speech datasets with dialect data, which supports the development of downstream applications such as speech generation. 

\newpage
\appendix
\section*{Appendix}

\section{Data Distribution}
\label{sec:distribution}

Attached is the training distribution for different dialects.

\begin{figure}[ht] {
    \centering
    \includegraphics[width=0.6\linewidth]{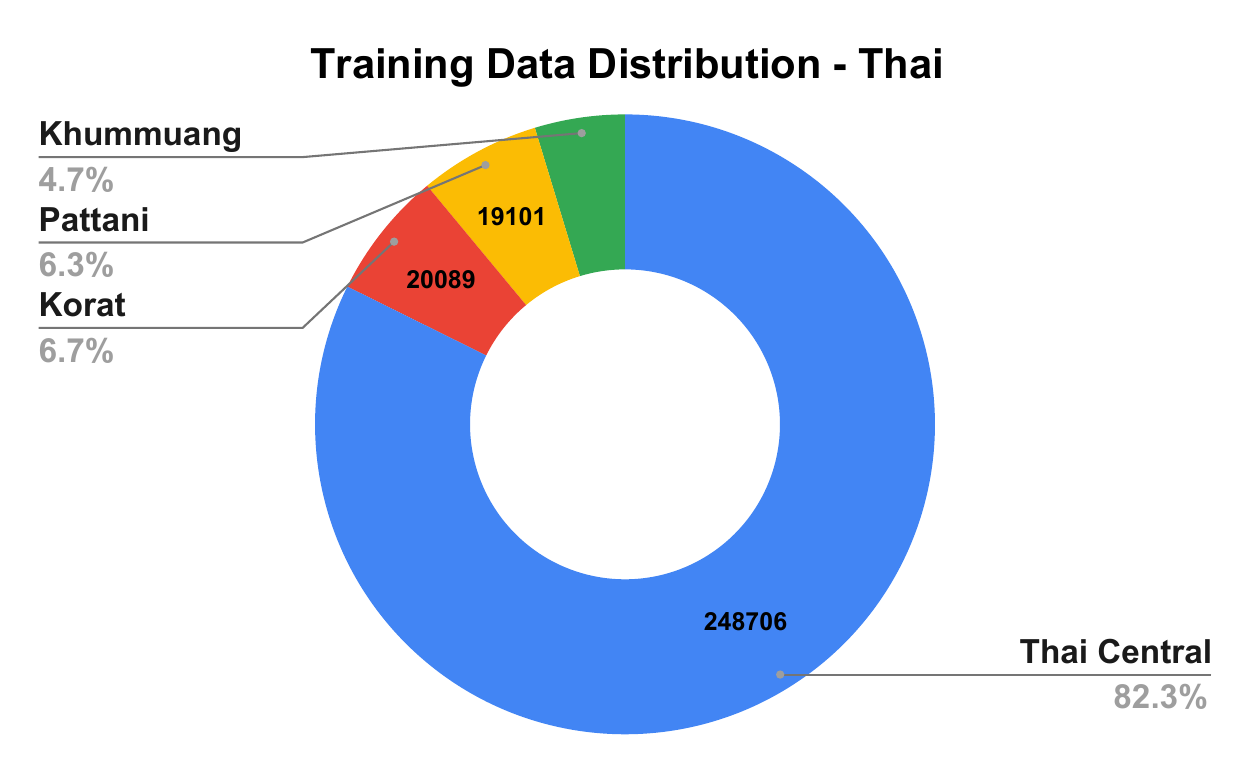}
    
    \caption{Training distribution for Thai dialect classification.}
    
    \label{fig:thai}
} \end{figure}

\begin{figure}[ht] {
    \centering
    \includegraphics[width=0.6\linewidth]{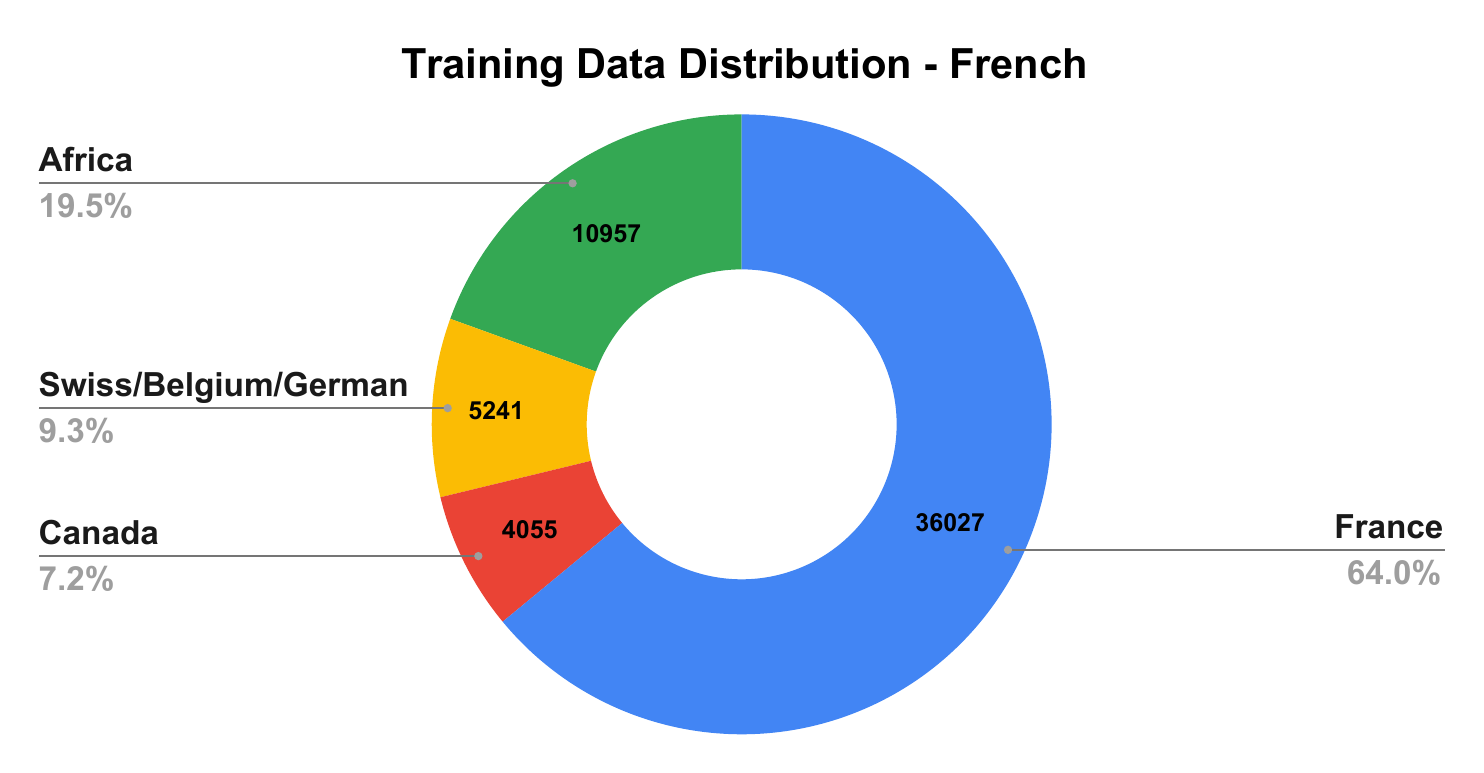}
    
    \caption{Training distribution for French dialect classification.}
    
    \label{fig:french}
} \end{figure}

\begin{figure}[ht] {
    \centering
    \includegraphics[width=0.6\linewidth]{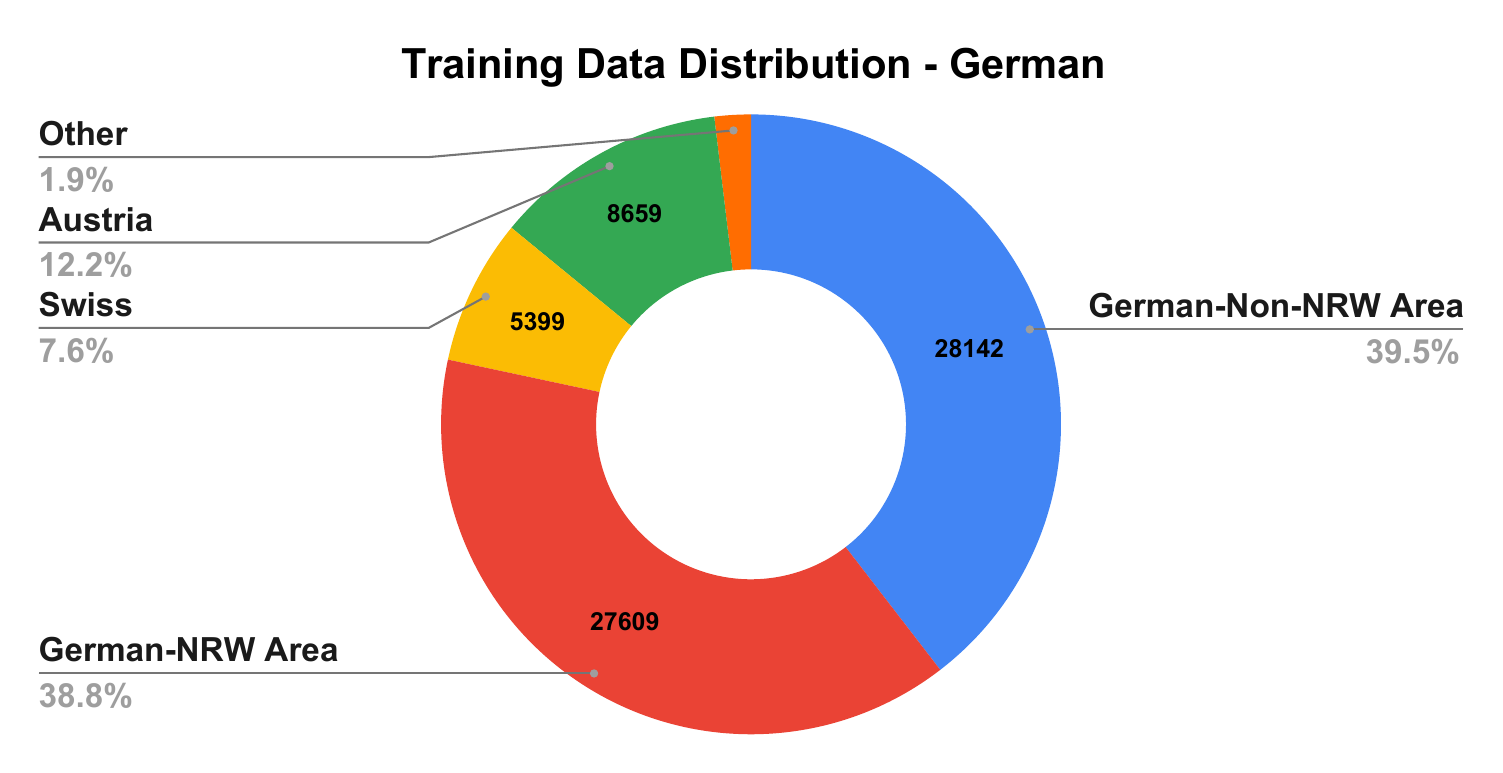}
    
    \caption{Training distribution for German dialect classification.}
    
    \label{fig:german}
} \end{figure}

\newpage

\begin{figure}[ht] {
    \centering
    \includegraphics[width=0.55\linewidth]{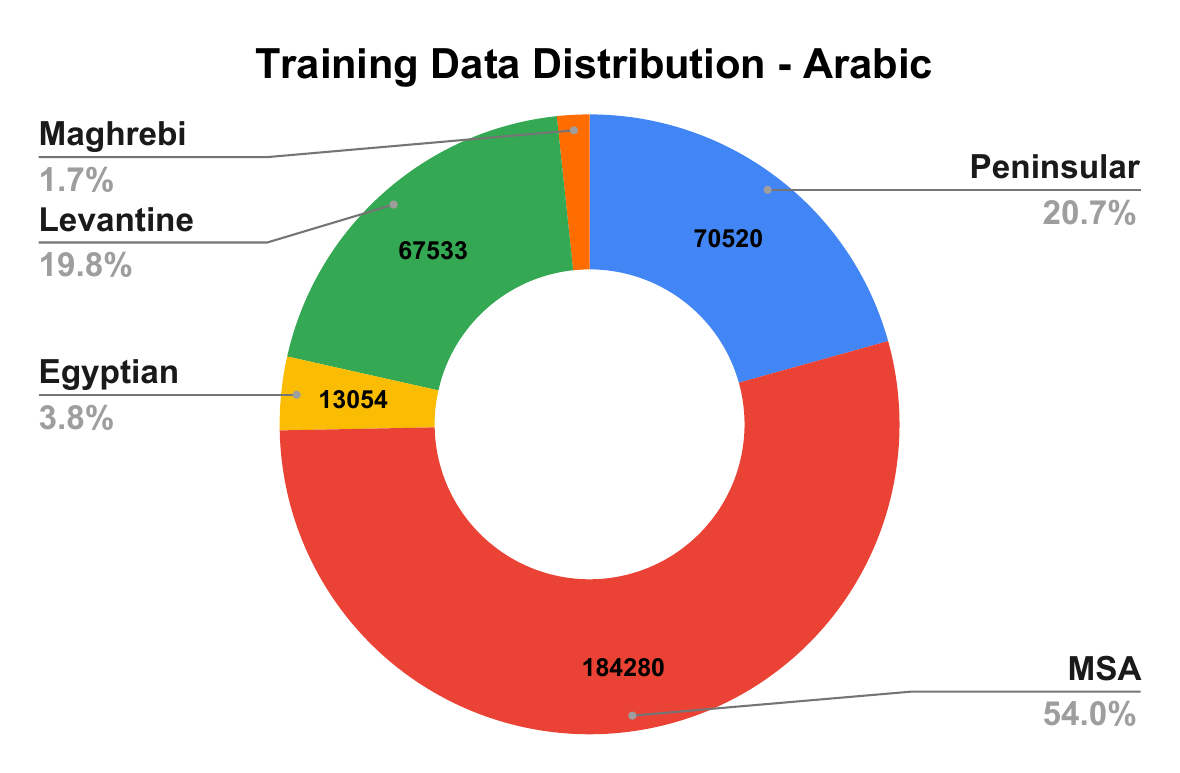}
    
    \caption{Training distribution for Arabic dialect classification.}
    
    \label{fig:arabic}
} \end{figure}

\begin{figure}[ht] {
    \centering
    \includegraphics[width=0.6\linewidth]{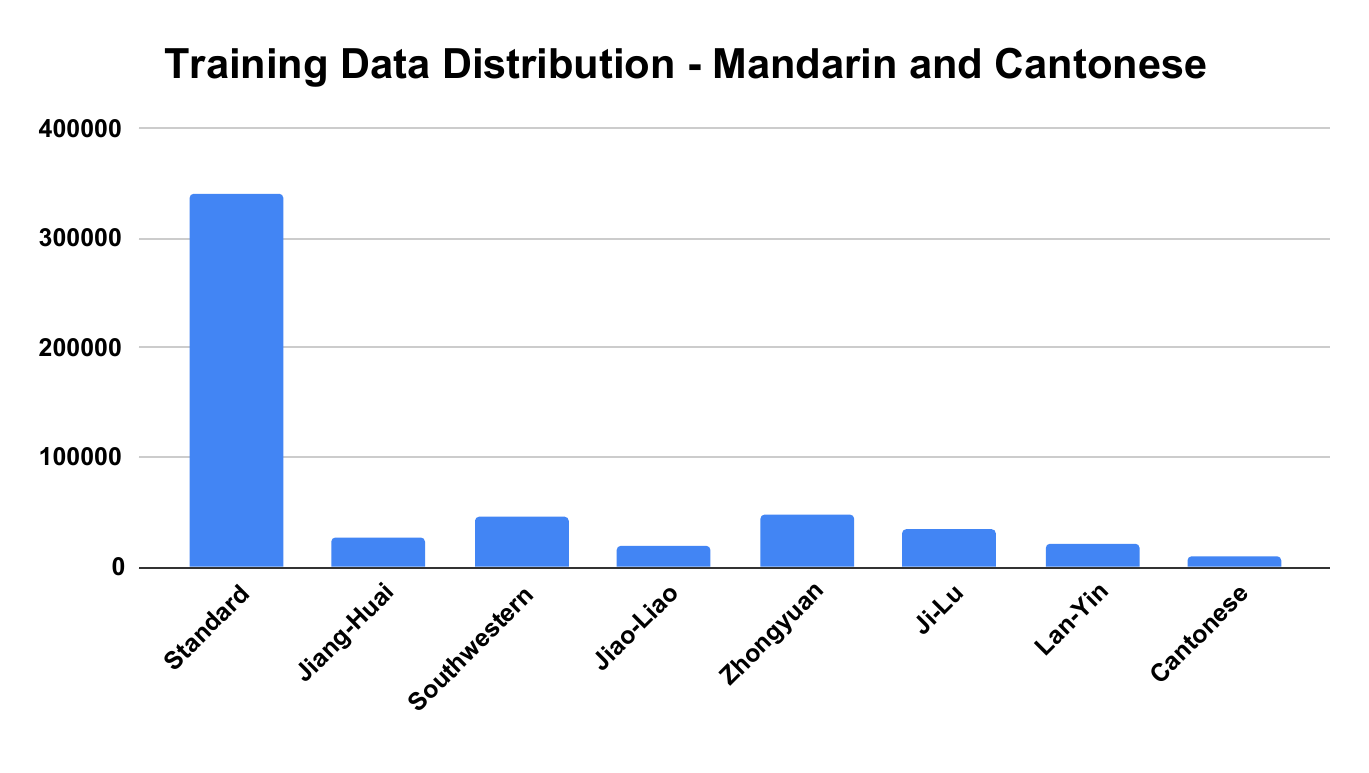}
    
    \caption{Training distribution for Mandarin dialect and Cantonese classification.}
    
    \label{fig:chinese}
} \end{figure}

\begin{figure}[ht] {
    \centering
    \includegraphics[width=0.5\linewidth]{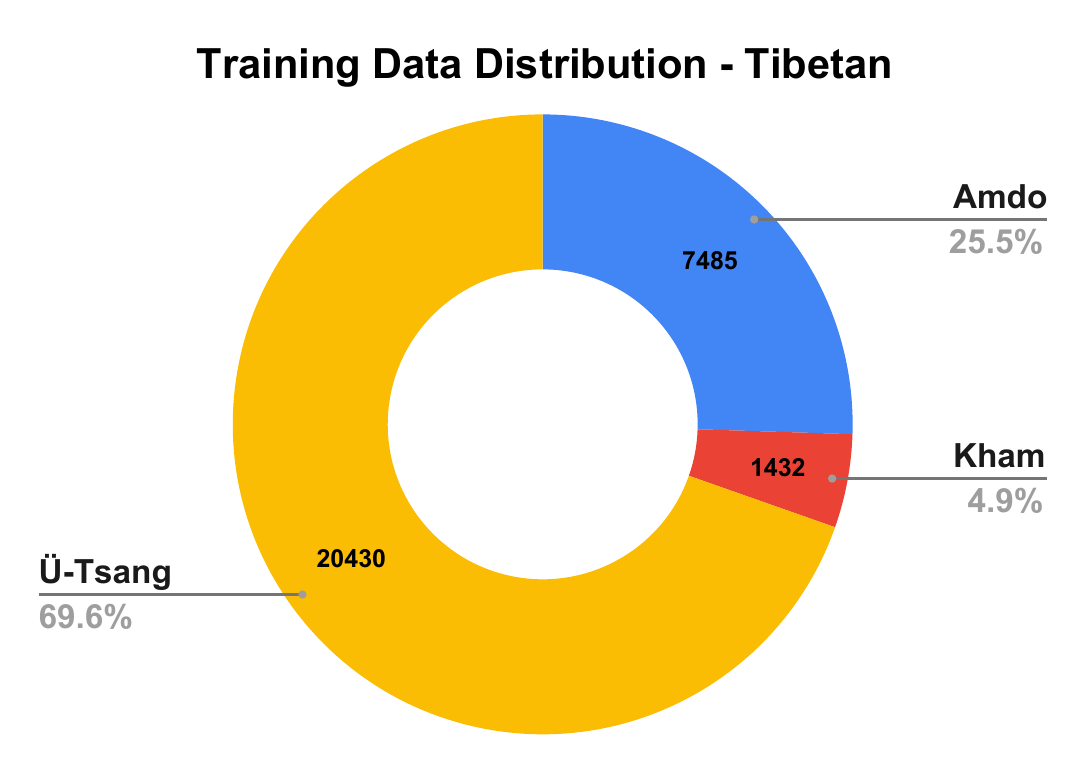}
    
    \caption{Training distribution for Tibetan (in China) dialect classification.}
    
    \label{fig:tibetan}
} \end{figure}

\newpage

\begin{figure}[ht] {
    \centering
    \includegraphics[width=0.8\linewidth]{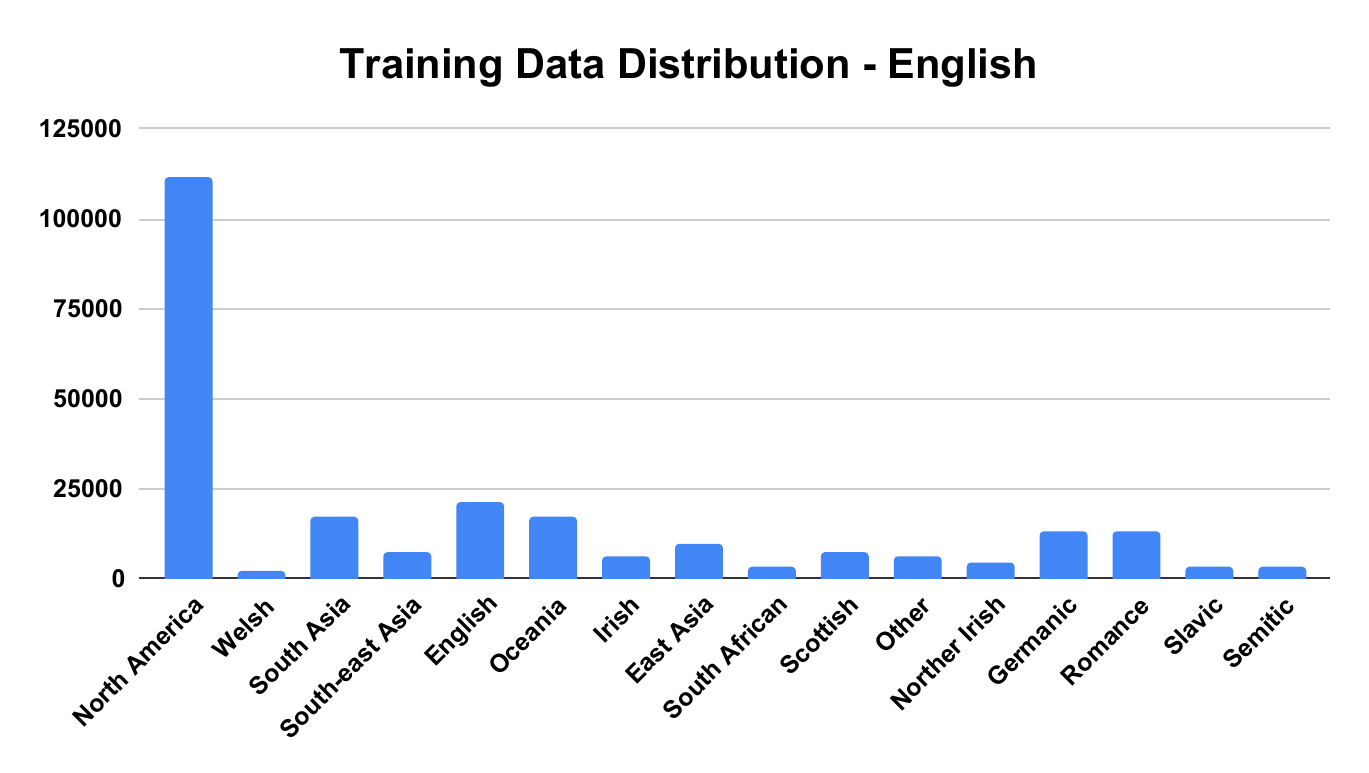}
    \vspace{-3mm}
    \caption{Training distribution for English dialect classification.}
    \vspace{-3mm}
    \label{fig:english}
} \end{figure}

\begin{figure}[ht] {
    \centering
    \includegraphics[width=0.6\linewidth]{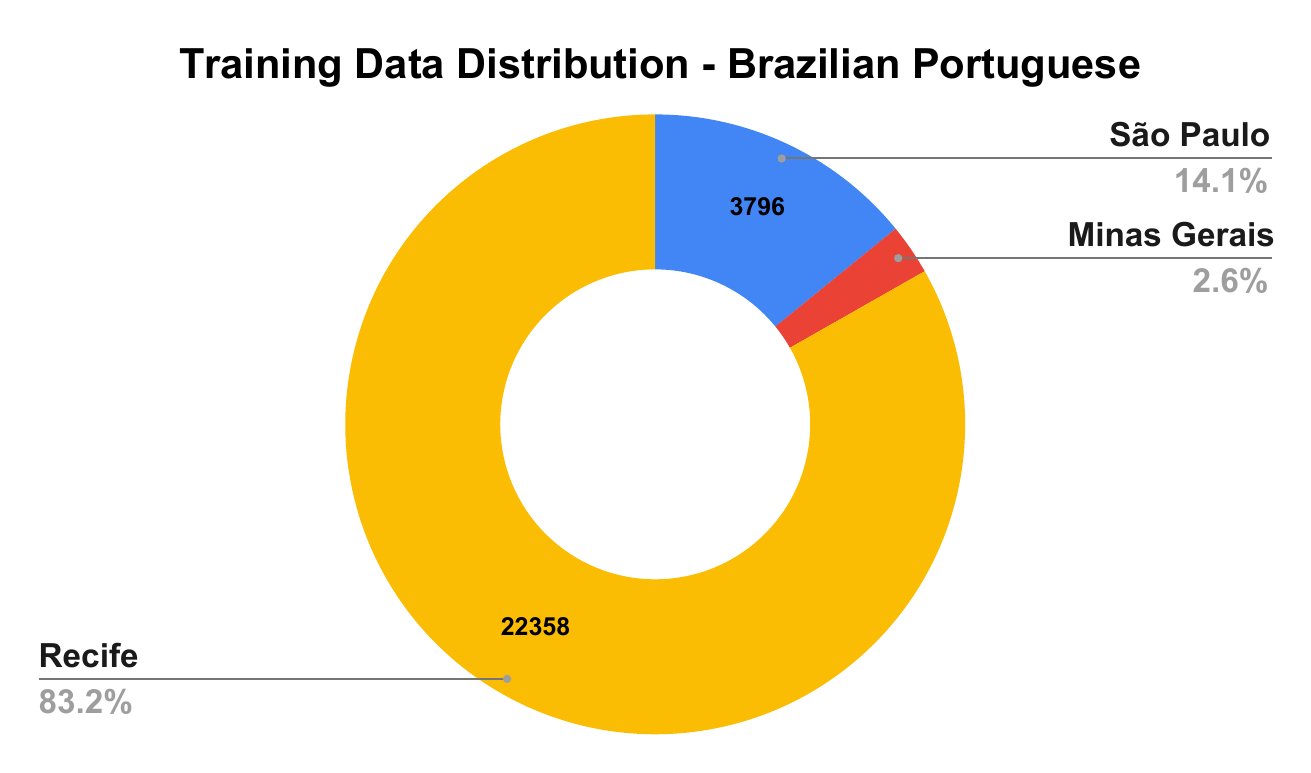}
    
    \caption{Training distribution for Brazilian Portuguese dialect classification.}
    \vspace{-3mm}
    \label{fig:brazil_portuguese}
} \end{figure}

\begin{figure}[ht] {
    \centering
    \includegraphics[width=0.6\linewidth]{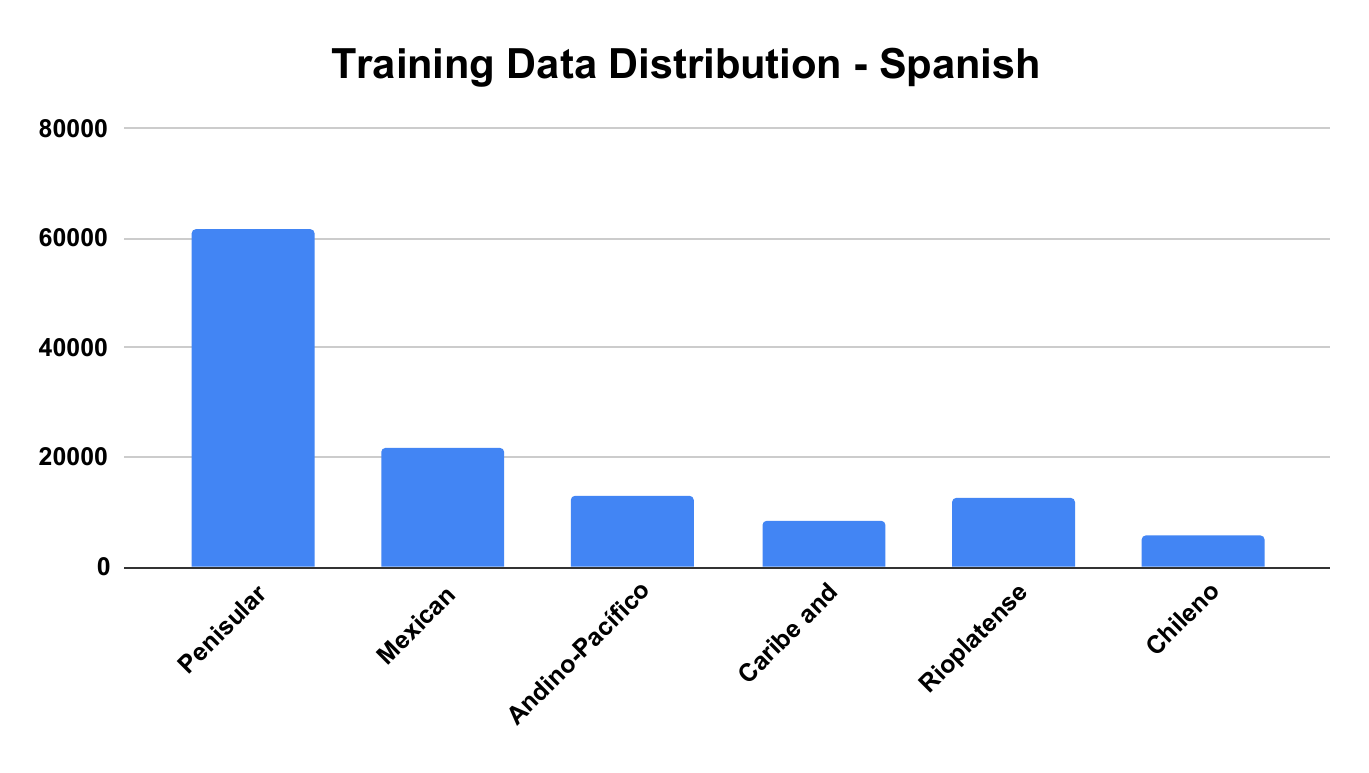}
    
    \caption{Training distribution for Spanish dialect classification.}
    
    \label{fig:spanish}
} \end{figure}

\newpage

\section{Confusion Matrix}
\label{sec:confusion_matrix}

\begin{figure}[ht] {
    \centering
    \includegraphics[width=0.4\linewidth]{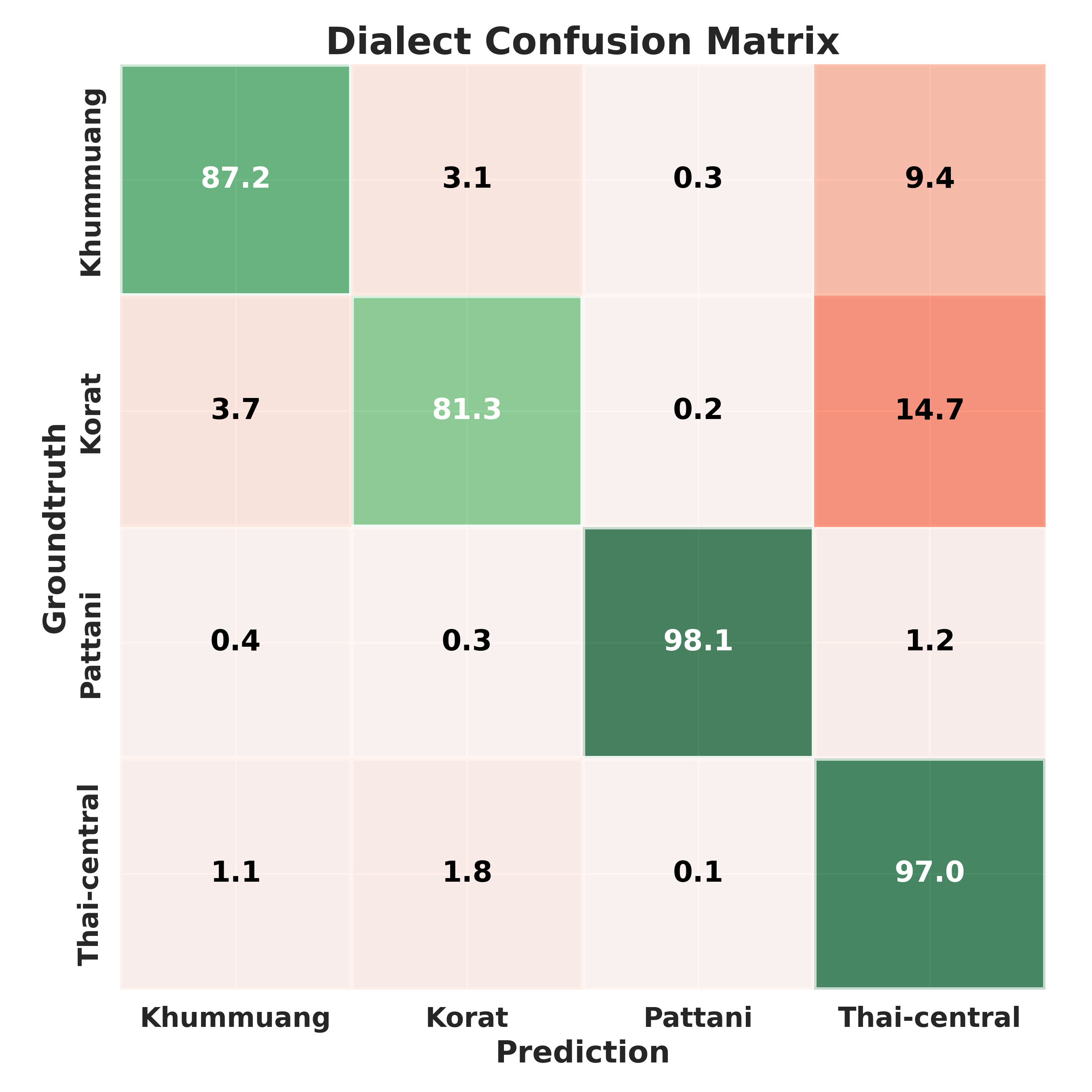}
    
    \caption{Confusion Matrix of Thai dialect prediction.}
    
    \label{fig:conf_thai}
} \end{figure}

\begin{figure}[ht] {
    \centering
    \includegraphics[width=0.4\linewidth]{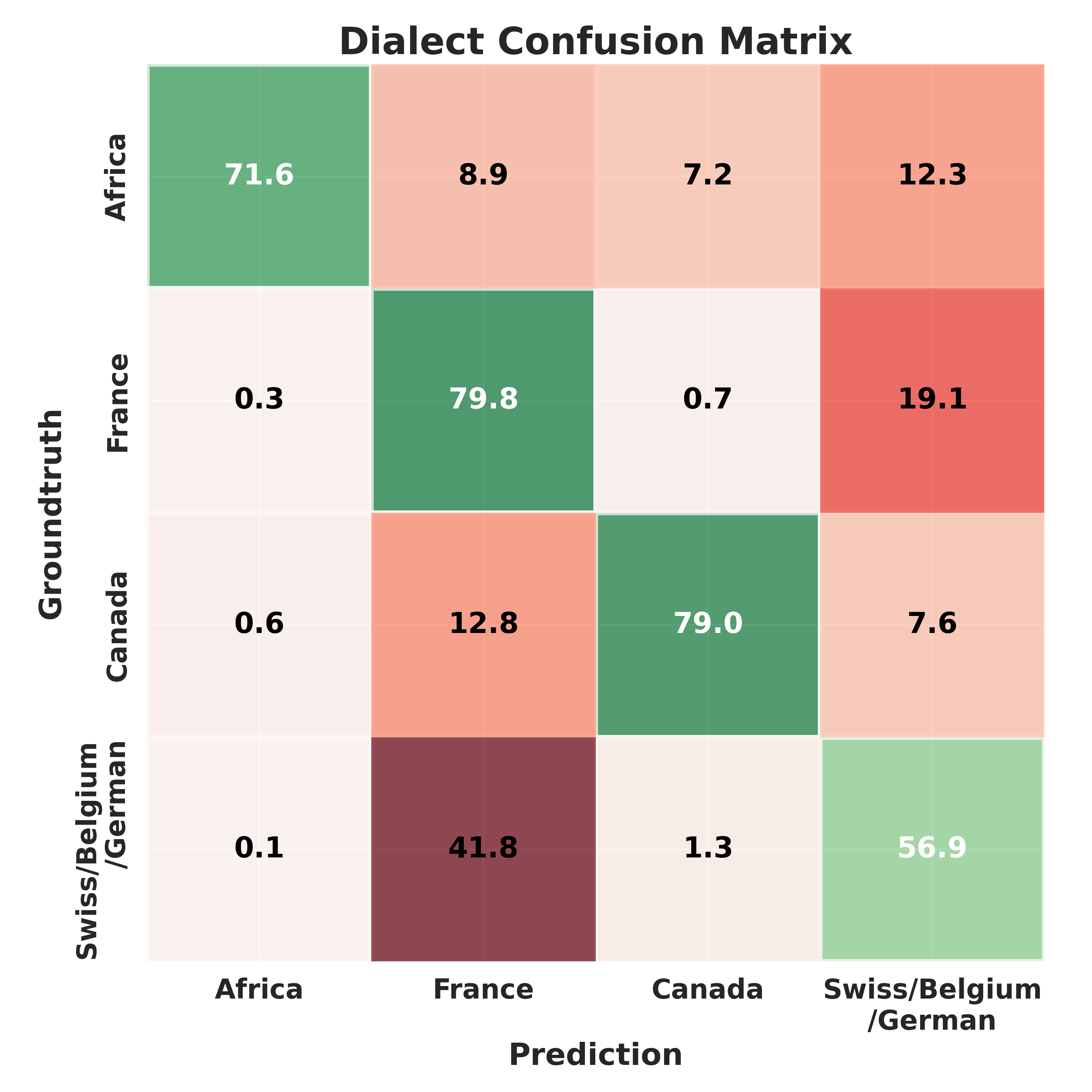}
    
    \caption{Confusion Matrix of French dialect prediction.}
    
    \label{fig:conf_french}
} \end{figure}

\begin{figure}[ht] {
    \centering
    \includegraphics[width=0.45\linewidth]{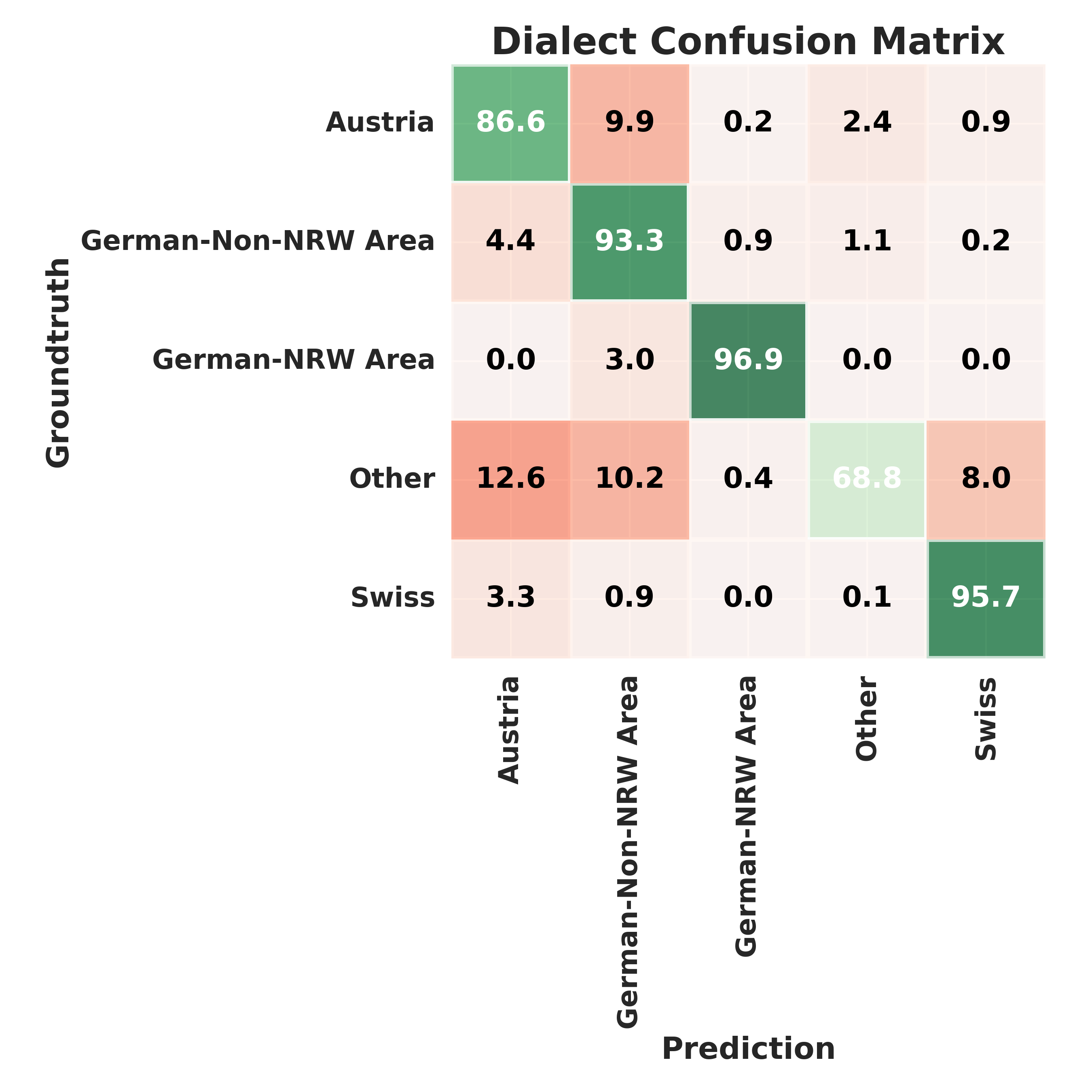}
    
    \caption{Confusion Matrix of German dialect prediction.}
    \vspace{-3.5mm}
    \label{fig:conf_german}
} \end{figure}

\begin{figure}[ht] {
    \centering
    \includegraphics[width=0.35\linewidth]{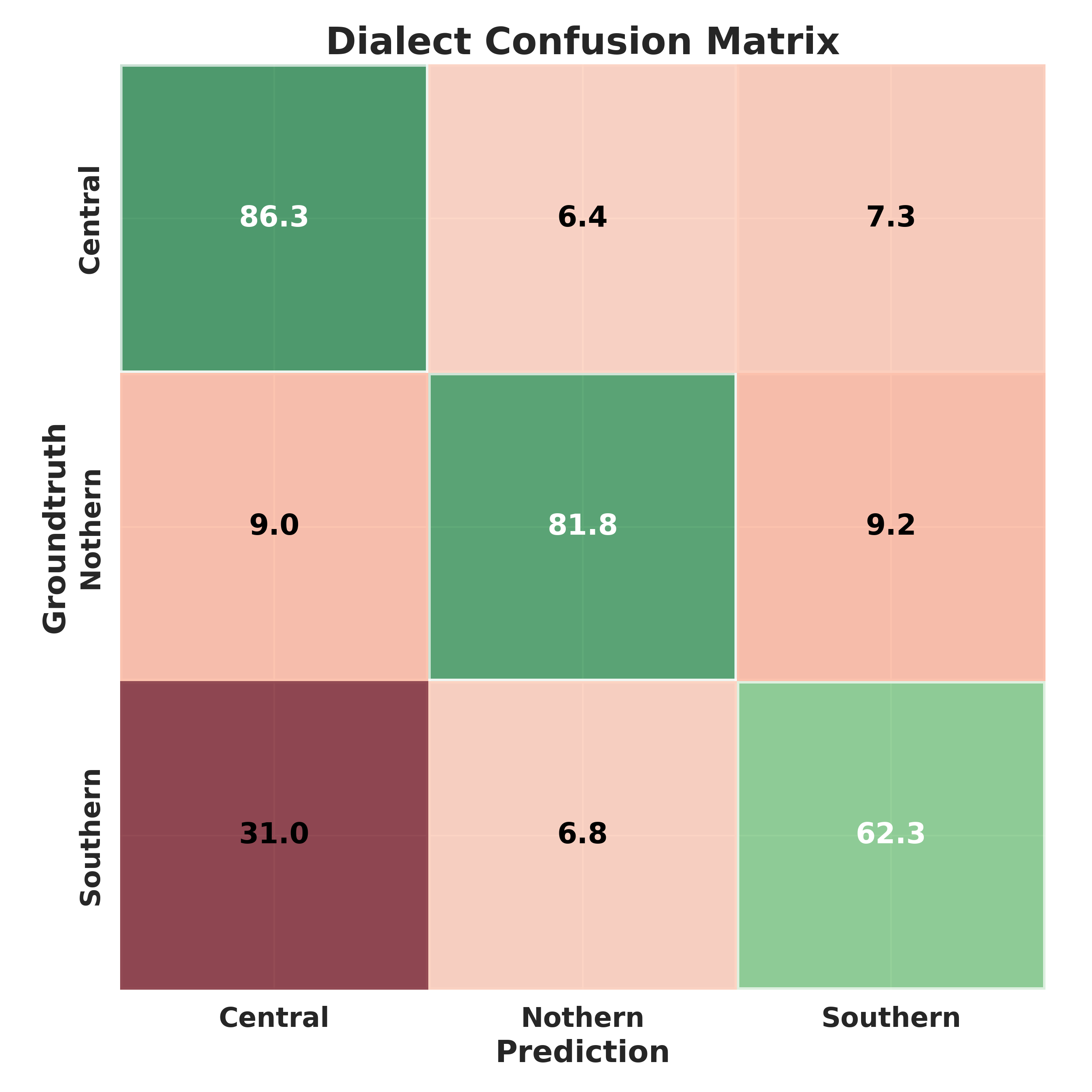}
    
    \caption{Confusion Matrix of Italian dialect prediction.}
    \vspace{-3.5mm}
    \label{fig:conf_italian}
} \end{figure}

\begin{figure}[ht] {
    \centering
    \includegraphics[width=0.35\linewidth]{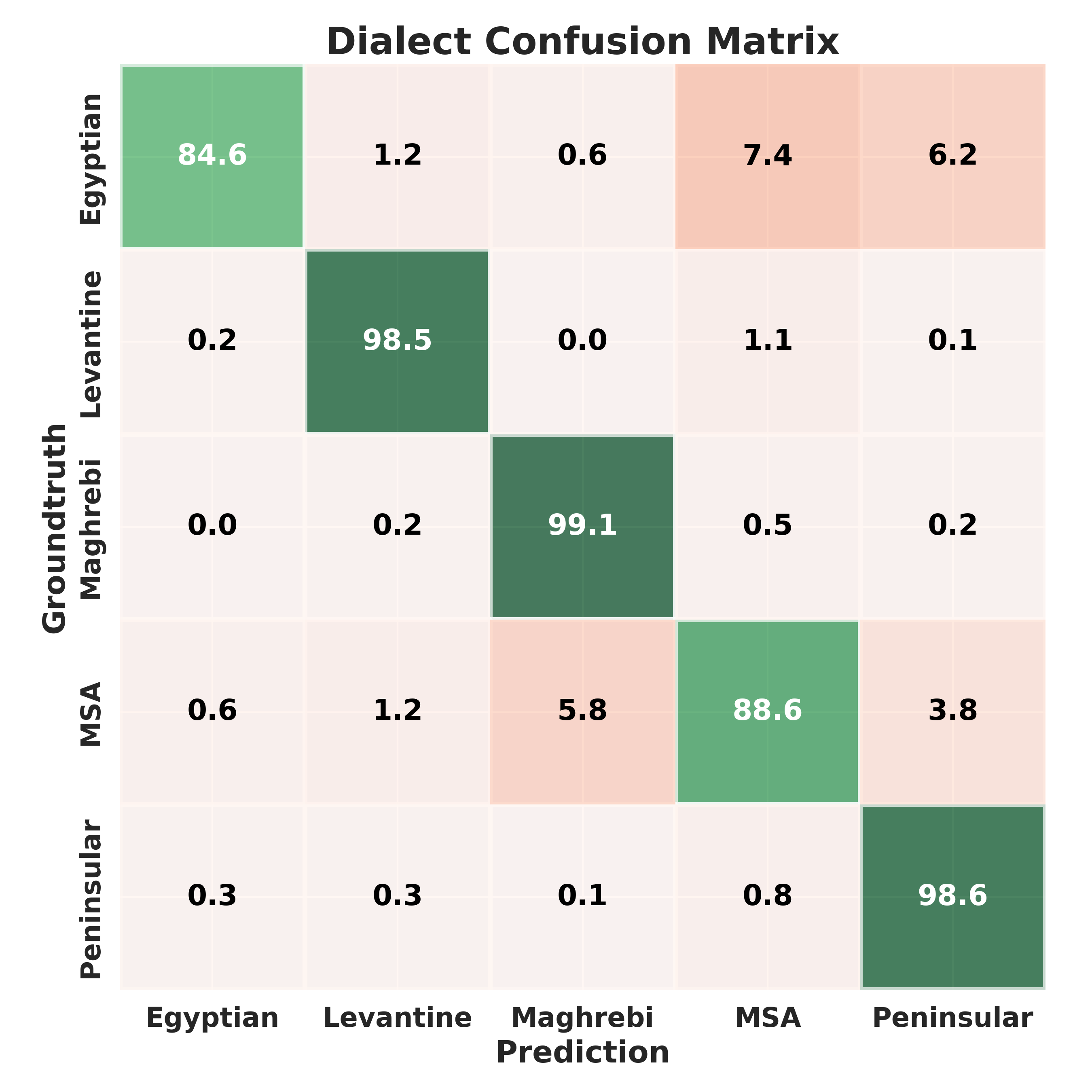}
    
    \caption{Confusion Matrix of Arabic dialect prediction.}
    \vspace{-3.5mm}
    \label{fig:conf_arabic}
} \end{figure}

\begin{figure}[ht] {
    \centering
    \includegraphics[width=\linewidth]{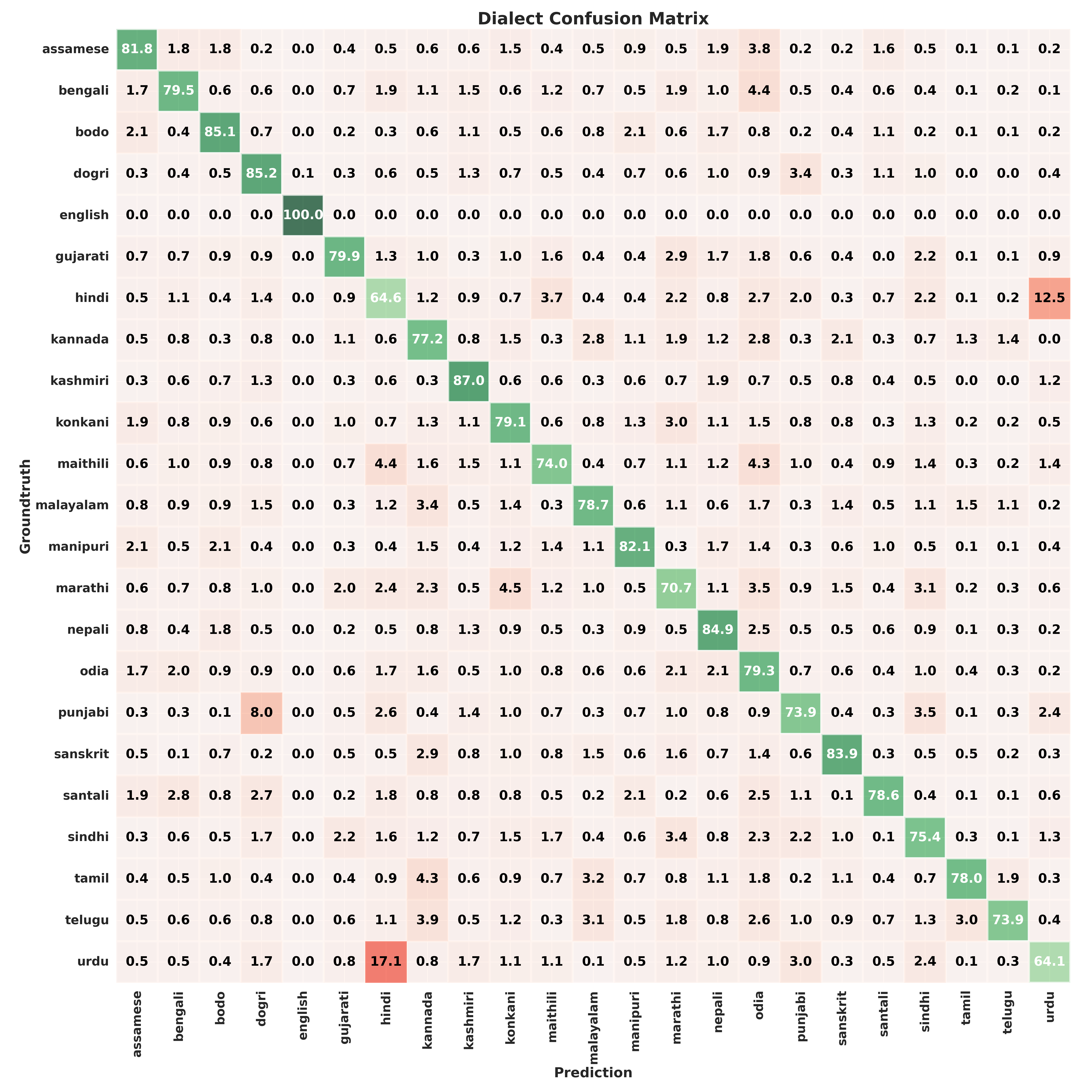}
    
    \caption{Confusion Matrix of Indic languages prediction.}
    
    \label{fig:conf_indic}
} \end{figure}

Confusion matrices for Thai, French, German, Italian, Arabic, and Indic languages are shown in Figure~\ref{fig:conf_thai}, ~\ref{fig:conf_french}, ~\ref{fig:conf_german}, ~\ref{fig:conf_italian}, ~\ref{fig:conf_arabic}, and ~\ref{fig:conf_indic}, respectively. 

\newpage

\newpage

\newpage

\section{Prompt}
\label{sec:prompt}

The text prompts for the speech generation of Mandarin speech are listed below:

1. \begin{CJK*}{UTF8}{gbsn}"早上，小兔子玉玉背着胡萝卜书包去上学。"\end{CJK*}

2. \begin{CJK*}{UTF8}{gbsn}“今天是个特别的日子,年度南方森林运动会。”\end{CJK*}

3. \begin{CJK*}{UTF8}{gbsn}“班主任熊猫翁大大说：“玉玉，你代表咱们班参加吧！””\end{CJK*}

4. \begin{CJK*}{UTF8}{gbsn}“有一次，北风和太阳正在争论谁比较有本事。”\end{CJK*}

5. \begin{CJK*}{UTF8}{gbsn}“他们正好看到有个人走过，那个人穿著一件斗篷。”\end{CJK*}

6. \begin{CJK*}{UTF8}{gbsn}“他们就说了，谁可以让那个人脱掉那件斗篷，就算谁比较有本事。”\end{CJK*}

7. \begin{CJK*}{UTF8}{gbsn}“于是，北风就拼命地吹。怎料，他吹得越厉害，那个人就越是用斗篷包紧自己。最后，北风没办法，只好放弃。”\end{CJK*}

8. \begin{CJK*}{UTF8}{gbsn}“接著，太阳出来晒了一下，那个人就立刻把斗篷脱掉了。于是，北风只好认输了。”\end{CJK*}

9. \begin{CJK*}{UTF8}{gbsn}“中国疆域广阔，人口众多，首都在北京。官方语言是普通话，使用汉字。同时，中华各民族儿女也使用民族语文。”\end{CJK*}

10. \begin{CJK*}{UTF8}{gbsn}“这里风光旖旎，众多山河湖海，凭借南北各种美景，吸引了众多游客。”\end{CJK*}

\newpage

\bibliographystyle{IEEEbib}
\bibliography{ref}

\begin{thebibliography}{10}

\bibitem{harris2024modeling}
Camille Harris, Chijioke Mgbahurike, Neha Kumar, and Diyi Yang,
\newblock ``Modeling gender and dialect bias in automatic speech recognition,''
\newblock in {\em Findings of the Association for Computational Linguistics: EMNLP 2024}, 2024, pp. 15166--15184.

\bibitem{chang2024self}
Kalvin Chang, Yi-Hui Chou, Jiatong Shi, Hsuan-Ming Chen, Nicole Holliday, Odette Scharenborg, and David~R Mortensen,
\newblock ``Self-supervised speech representations still struggle with african american vernacular english,''
\newblock in {\em Proc. Interspeech 2024}, 2024, pp. 4643--4647.

\bibitem{tang2021kespeech}
Zhiyuan Tang, Dong Wang, Yanguang Xu, Jianwei Sun, Xiaoning Lei, Shuaijiang Zhao, Cheng Wen, Xingjun Tan, Chuandong Xie, Shuran Zhou, et~al.,
\newblock ``Kespeech: An open source speech dataset of mandarin and its eight subdialects,''
\newblock in {\em Thirty-fifth Conference on Neural Information Processing Systems Datasets and Benchmarks Track (Round 2)}, 2021.

\bibitem{sanabria2023edinburgh}
Ramon Sanabria, Nikolay Bogoychev, Nina Markl, Andrea Carmantini, Ondrej Klejch, and Peter Bell,
\newblock ``The edinburgh international accents of english corpus: Towards the democratization of english asr,''
\newblock in {\em ICASSP 2023-2023 IEEE International Conference on Acoustics, Speech and Signal Processing (ICASSP)}. IEEE, 2023, pp. 1--5.

\bibitem{demirsahin2020open}
Isin Demirsahin, Oddur Kjartansson, Alexander Gutkin, and Clara Rivera,
\newblock ``Open-source multi-speaker corpora of the english accents in the british isles,''
\newblock in {\em Proceedings of the twelfth language resources and evaluation conference}, 2020, pp. 6532--6541.

\bibitem{pratap2024scaling}
Vineel Pratap, Andros Tjandra, Bowen Shi, Paden Tomasello, Arun Babu, Sayani Kundu, Ali Elkahky, Zhaoheng Ni, Apoorv Vyas, Maryam Fazel-Zarandi, et~al.,
\newblock ``Scaling speech technology to 1,000+ languages,''
\newblock {\em Journal of Machine Learning Research}, vol. 25, no. 97, pp. 1--52, 2024.

\bibitem{radford2023robust}
Alec Radford, Jong~Wook Kim, Tao Xu, Greg Brockman, Christine McLeavey, and Ilya Sutskever,
\newblock ``Robust speech recognition via large-scale weak supervision,''
\newblock in {\em International Conference on Machine Learning}. PMLR, 2023, pp. 28492--28518.

\bibitem{ardila2020common}
Rosana Ardila, Megan Branson, Kelly Davis, Michael Kohler, Josh Meyer, Michael Henretty, Reuben Morais, Lindsay Saunders, Francis Tyers, and Gregor Weber,
\newblock ``Common voice: A massively-multilingual speech corpus,''
\newblock in {\em Proceedings of the Twelfth Language Resources and Evaluation Conference}, 2020, pp. 4218--4222.

\bibitem{zuluaga2023commonaccent}
Juan Zuluaga-Gomez, Sara Ahmed, Danielius Visockas, and Cem Subakan,
\newblock ``Commonaccent: Exploring large acoustic pretrained models for accent classification based on common voice,''
\newblock in {\em Proc. Interspeech 2023}, 2023, pp. 5291--5295.

\bibitem{wang2024globe}
Wenbin Wang, Yang Song, and Sanjay Jha,
\newblock ``Globe: A high-quality english corpus with global accents for zero-shot speaker adaptive text-to-speech,''
\newblock in {\em Proc. Interspeech 2024}, 2024, pp. 1365--1369.

\bibitem{diwan2025scaling}
Anuj Diwan, Zhisheng Zheng, David Harwath, and Eunsol Choi,
\newblock ``Scaling rich style-prompted text-to-speech datasets,''
\newblock {\em arXiv preprint arXiv:2503.04713}, 2025.

\bibitem{feng2025vox}
Tiantian Feng, Jihwan Lee, Anfeng Xu, Yoonjeong Lee, Thanathai Lertpetchpun, Xuan Shi, Helin Wang, Thomas Thebaud, Laureano Moro-Velazquez, Dani Byrd, et~al.,
\newblock ``Vox-profile: A speech foundation model benchmark for characterizing diverse speaker and speech traits,''
\newblock {\em arXiv preprint arXiv:2505.14648}, 2025.

\bibitem{shi2021aishell}
Yao Shi, Hui Bu, Xin Xu, Shaoji Zhang, and Ming Li,
\newblock ``Aishell-3: A multi-speaker mandarin tts corpus,''
\newblock in {\em Proc. Interspeech 2021}, 2021, pp. 2756--2760.

\bibitem{sullivan2023robustness}
Peter Sullivan, AbdelRahim Elmadany, and Muhammad Abdul-Mageed,
\newblock ``On the robustness of arabic speech dialect identification,''
\newblock in {\em Proc. Interspeech 2023}, 2023, pp. 5326--5330.

\bibitem{chen2022wavlm}
Sanyuan Chen, Chengyi Wang, Zhengyang Chen, Yu~Wu, Shujie Liu, Zhuo Chen, Jinyu Li, Naoyuki Kanda, Takuya Yoshioka, Xiong Xiao, et~al.,
\newblock ``Wavlm: Large-scale self-supervised pre-training for full stack speech processing,''
\newblock {\em IEEE Journal of Selected Topics in Signal Processing}, vol. 16, no. 6, pp. 1505--1518, 2022.

\bibitem{gadm2024}
{GADM},
\newblock ``Database of global administrative areas, version 4.1,'' \url{https://gadm.org/}, 2025,
\newblock Accessed on 28 July 2025.

\bibitem{lander2005cslu}
T~Lander,
\newblock ``Cslu: 22 languages corpus (ldc2005s26),''
\newblock {\em Linguistic Data Consortium}, 2005.

\bibitem{zhao2018l2}
Guanlong Zhao, Sinem Sonsaat, Alif Silpachai, Ivana Lucic, Evgeny Chukharev-Hudilainen, John Levis, and Ricardo Gutierrez-Osuna,
\newblock ``L2-arctic: A non-native english speech corpus,''
\newblock in {\em Proc. Interspeech 2018}, 2018, pp. 2783--2787.

\bibitem{garofolo1993darpa}
John~S Garofolo, Lori~F Lamel, William~M Fisher, Jonathan~G Fiscus, and David~S Pallett,
\newblock ``Darpa timit acoustic-phonetic continous speech corpus cd-rom. nist speech disc 1-1.1,''
\newblock {\em NASA STI/Recon technical report n}, vol. 93, pp. 27403, 1993.

\bibitem{wang2021voxpopuli}
Changhan Wang, Morgane Riviere, Ann Lee, Anne Wu, Chaitanya Talnikar, Daniel Haziza, Mary Williamson, Juan Pino, and Emmanuel Dupoux,
\newblock ``Voxpopuli: A large-scale multilingual speech corpus for representation learning, semi-supervised learning and interpretation,''
\newblock in {\em Proceedings of the 59th Annual Meeting of the Association for Computational Linguistics and the 11th International Joint Conference on Natural Language Processing (Volume 1: Long Papers)}, 2021, pp. 993--1003.

\bibitem{wang2024usat}
Wenbin Wang, Yang Song, and Sanjay Jha,
\newblock ``Usat: A universal speaker-adaptive text-to-speech approach,''
\newblock {\em IEEE/ACM Transactions on Audio, Speech, and Language Processing}, 2024.

\bibitem{veliche2024towards}
Irina-Elena Veliche, Zhuangqun Huang, Vineeth Ayyat~Kochaniyan, Fuchun Peng, Ozlem Kalinli, and Michael~L Seltzer,
\newblock ``Towards measuring fairness in speech recognition: Fair-speech dataset,''
\newblock in {\em Proc. Interspeech 2024}, 2024, pp. 1385--1389.

\bibitem{nigerian_eng}
``Crowdsourced high-quality nigerian english speech data set.,''
\newblock {\em Open Speech and Language Resources}, 2019.

\bibitem{hispanic_eng}
William Byrne, Eva Knodt, Jared Bernstein, and Farzhad Emami,
\newblock ``Hispanic-english database (ldc2014s05),''
\newblock {\em Linguistic Data Consortium}, 2014.

\bibitem{al2023masc}
Mohammad Al-Fetyani, Muhammad Al-Barham, Gheith Abandah, Adham Alsharkawi, and Maha Dawas,
\newblock ``Masc: Massive arabic speech corpus,''
\newblock in {\em 2022 IEEE Spoken Language Technology Workshop (SLT)}. IEEE, 2023, pp. 1006--1013.

\bibitem{alharbi2024sada}
Sadeen Alharbi, Areeb Alowisheq, Zolt{\'a}n T{\"u}ske, Kareem Darwish, Abdullah Alrajeh, Abdulmajeed Alrowithi, Aljawharah~Bin Tamran, Asma Ibrahim, Raghad Aloraini, Raneem Alnajim, et~al.,
\newblock ``Sada: Saudi audio dataset for arabic,''
\newblock in {\em ICASSP 2024-2024 IEEE International Conference on Acoustics, Speech and Signal Processing (ICASSP)}. IEEE, 2024, pp. 10286--10290.

\bibitem{benelallamdvoice}
I~Benelallam, AM~Naira, and A~Allak,
\newblock ``Dvoice: an open source dataset for automatic speech recognition on moroccan dialectal arabic (2021),'' .

\bibitem{zhao2020open}
Yue Zhao, Xiaona Xu, Jianjian Yue, Wei Song, Xiali Li, Licheng Wu, and Qiang Ji,
\newblock ``An open speech resource for tibetan multi-dialect and multitask recognition,''
\newblock {\em International Journal of Computational Science and Engineering}, vol. 22, no. 2-3, pp. 297--304, 2020.

\bibitem{javed2024indicvoices}
Tahir Javed, Janki Nawale, Eldho George, Sakshi Joshi, Kaushal Bhogale, Deovrat Mehendale, Ishvinder Sethi, Aparna Ananthanarayanan, Hafsah Faquih, Pratiti Palit, et~al.,
\newblock ``Indicvoices: Towards building an inclusive multilingual speech dataset for indian languages,''
\newblock in {\em Findings of the Association for Computational Linguistics ACL 2024}, 2024, pp. 10740--10782.

\bibitem{suwanbandit2023thai}
Artit Suwanbandit, Burin Naowarat, Orathai Sangpetch, and Ekapol Chuangsuwanich,
\newblock ``Thai dialect corpus and transfer-based curriculum learning investigation for dialect automatic speech recognition,''
\newblock in {\em Proc. Interspeech}, 2023, vol.~2.

\bibitem{guevara2020crowdsourcing}
Adriana Guevara-Rukoz, Isin Demirsahin, Fei He, Shan-Hui~Cathy Chu, Supheakmungkol Sarin, Knot Pipatsrisawat, Alexander Gutkin, Alena Butryna, and Oddur Kjartansson,
\newblock ``Crowdsourcing latin american spanish for low-resource text-to-speech,''
\newblock in {\em Proceedings of the Twelfth Language Resources and Evaluation Conference}, 2020, pp. 6504--6513.

\bibitem{koudounas2023italic}
Alkis Koudounas, Moreno La~Quatra, Lorenzo Vaiani, Luca Colomba, Giuseppe Attanasio, Eliana Pastor, Luca Cagliero, and Elena Baralis,
\newblock ``Italic: An italian intent classification dataset,''
\newblock in {\em Proc. Interspeech 2023}, 2023, pp. 2153--2157.

\bibitem{candido2023coraa}
Arnaldo Candido~Junior, Edresson Casanova, Anderson Soares, Frederico~Santos de~Oliveira, Lucas Oliveira, Ricardo Corso~Fernandes Junior, Daniel Peixoto~Pinto da~Silva, Fernando~Gorgulho Fayet, Bruno~Baldissera Carlotto, Lucas Rafael~Stefanel Gris, et~al.,
\newblock ``Coraa asr: a large corpus of spontaneous and prepared speech manually validated for speech recognition in brazilian portuguese,''
\newblock {\em Language Resources and Evaluation}, vol. 57, no. 3, pp. 1139--1171, 2023.

\bibitem{zhong2025accentbox}
Jinzuomu Zhong, Korin Richmond, Zhiba Su, and Siqi Sun,
\newblock ``Accentbox: Towards high-fidelity zero-shot accent generation,''
\newblock in {\em ICASSP 2025-2025 IEEE International Conference on Acoustics, Speech and Signal Processing (ICASSP)}. IEEE, 2025, pp. 1--5.

\bibitem{resnick2012phonological}
Melvyn~C Resnick,
\newblock {\em Phonological variants and dialect identification in Latin American Spanish}, vol. 201,
\newblock Walter de Gruyter, 2012.

\bibitem{hu2024wavllm}
Shujie Hu, Long Zhou, Shujie Liu, Sanyuan Chen, Lingwei Meng, Hongkun Hao, Jing Pan, Xunying Liu, Jinyu Li, Sunit Sivasankaran, et~al.,
\newblock ``Wavllm: Towards robust and adaptive speech large language model,''
\newblock {\em arXiv preprint arXiv:2404.00656}, 2024.

\bibitem{pepino2021emotion}
Leonardo Pepino, Pablo Riera, and Luciana Ferrer,
\newblock ``Emotion recognition from speech using wav2vec 2.0 embeddings,''
\newblock in {\em Proc. Interspeech 2021}, 2021, pp. 3400--3404.

\bibitem{hu2022lora}
Edward~J Hu, Yelong Shen, Phillip Wallis, Zeyuan Allen-Zhu, Yuanzhi Li, Shean Wang, Lu~Wang, Weizhu Chen, et~al.,
\newblock ``Lora: Low-rank adaptation of large language models.,''
\newblock {\em ICLR}, vol. 1, no. 2, pp. 3, 2022.

\bibitem{feng2023peft}
Tiantian Feng and Shrikanth Narayanan,
\newblock ``Peft-ser: On the use of parameter efficient transfer learning approaches for speech emotion recognition using pre-trained speech models,''
\newblock in {\em 2023 11th International Conference on Affective Computing and Intelligent Interaction (ACII)}. IEEE, 2023, pp. 1--8.

\bibitem{chambers1998dialectology}
Jack~K Chambers and Peter Trudgill,
\newblock {\em Dialectology},
\newblock Cambridge University Press, 1998.

\bibitem{nerbonne2010measuring}
John Nerbonne,
\newblock ``Measuring the diffusion of linguistic change,''
\newblock {\em Philosophical Transactions of the Royal Society B: Biological Sciences}, vol. 365, no. 1559, pp. 3821--3828, 2010.

\bibitem{du2024cosyvoice}
Zhihao Du, Yuxuan Wang, Qian Chen, Xian Shi, Xiang Lv, Tianyu Zhao, Zhifu Gao, Yexin Yang, Changfeng Gao, Hui Wang, et~al.,
\newblock ``Cosyvoice 2: Scalable streaming speech synthesis with large language models,''
\newblock {\em arXiv preprint arXiv:2412.10117}, 2024.

\end{thebibliography}

\end{document}